\documentclass[pdflatex, preprint, bst/sn-mathphys-num]{sn-jnl}
\usepackage{graphicx}%
\usepackage{multirow}%
\usepackage{amsmath,amssymb,amsfonts}%
\usepackage{amsthm}%
\usepackage{mathrsfs}%
\usepackage[title]{appendix}%
\usepackage{xcolor}%
\usepackage{textcomp}%
\usepackage{manyfoot}%
\usepackage{booktabs}%
\usepackage[numbers, sort&compress]{natbib}
\usepackage{overpic}

\newcommand{\Aarnes}{\citet{aarnes2025}}
\newcommand{\Babiker}{{\citet{babiker2023}}}

\renewcommand{\rm}[1]{\mathrm{#1}}
\newcommand{\Taylor}{\lambda}

\newcommand{\Lint}{L_\infty}
\newcommand{\Tint}{T_\infty}
\newcommand{\Frt}{\rm{Fr}_{\rm{T}}}
\newcommand{\Wet}{\rm{We}_{\rm{T}}}
\newcommand{\ReynoldsNumber}{\rm{Re}}
\newcommand{\Reint}{\ReynoldsNumber_\rm{T}}
\newcommand{\Rel}{\ReynoldsNumber_\lambda}

\newcommand{\meanT}[1]{\langle #1 \rangle}
\newcommand{\meanH}[1]{\overline{#1}}
\newcommand{\mean}[1]{\meanT{\meanH{#1}}}

\newcommand{\MSbetas}{\meanH{\beta_s^2}}

\newcommand{\MSphi}{\meanH{\phi^2}}

\newcommand{\sigfi}{\sigma_\sfield^2}
\newcommand{\sigbeta}{\sigma_\beta^2}
\newcommand{\sigom}{\sigma_\omega^2}

\newcommand{\sfield}{\phi}

\newcommand{\lag}{\tau}

\newcommand{\obsWindow}{W}
\newcommand{\subWindow}{A}
\newcommand{\intensity}{\nu}

\newcommand{\omz}{\omega}
\newcommand{\subs}{_{(s)}}
\newcommand{\subd}{_{(d)}}
\newcommand{\lagmin}{\lag_{\min}}
\newcommand{\lagmax}{\lag_{\max}}

\begin{document}

\title[On the spatial statistics of free-surface turbulence and the complementarity of `dimples’ and `scars’]{On the spatial statistics of free-surface turbulence and the complementarity of `dimples’ and `scars’}

\author*[1]{\fnm{Daniel R.} \sur{Kjellevold}}\email{daniel.kjellevold@ntnu.no}

\author[2]{\fnm{J{\o}rgen R.} \sur{Aarnes}}\email{jorgen.r.aarnes@ntnu.no}

\author[2]{\fnm{Omer M.} \sur{Babiker}}\email{omer.babiker@ntnu.no}

\author[1]{\fnm{Ingelin} \sur{Steinsland}}\email{ingelin.steinsland@ntnu.no}

\author[2]{\fnm{Simen {\AA}.} \sur{Ellingsen}}\email{simen.a.ellingsen@ntnu.no}

\affil*[1]{\orgdiv{Department of Mathematical Sciences}, \orgname{Norwegian University of Science and Technology}, \postcode{7491}, \city{Trondheim}, \country{Norway}}

\affil[2]{\orgdiv{Department of Energy and Process Engineering}, \orgname{Norwegian University of Science and Technology}, \postcode{7491}, \city{Trondheim}, \country{Norway}}

\abstract{The air--water interface governs the exchange of heat and gas between natural water bodies and their surrounding environment. Turbulence beneath the free surface imprints characteristic features: near-circular depressions (`dimples') and elongated indentations (`scars'). Recent studies have shown that these are temporally linked to sub-surface flow features, e.g., a rapid increases in mean-square surface divergence due to upwelling events, precede dimple count surges. Yet the spatial structure of these connections remains unquantified. We employ spatial statistics to consider the spatial and temporal correlations between dimples and scars and two key velocity-derived fields, surface divergence $\beta=\partial_x u+\partial_y v$ and vertical vorticity $\omega=\partial_x v-\partial_y u$. The dimples and scars are modelled as inhomogeneous Poisson point-processes, with intensity fields driven by the local variance of $\beta$ and $\omega$. Parameters are estimated by maximum likelihood against six DNS datasets, quantifying the spatial and temporal connection between dimples, scars, surface divergence and vorticity. Our results demonstrate a clear complementarity: Dimples show strong local connection to the vertical vorticity field but has weak spatial connection with surface divergence and a spatially ``global'' model is required for dimples to work as estimators of surface divergence; scars, in a similar but opposite manner, couple locally to surface divergence but globally to the vertical vorticity. The global picture of dimples is nuanced by separating the divergence contribution into positive and negative fields, which reveals how dimples are pushed away from upwelling regions and collect near downwelling regions.}




\maketitle

\section{Introduction}
\label{sec:intro}

The surface of a gently flowing river is a fascinating sight as it displays a multitude of features on the surface, imprints of the turbulent flow underneath it. Free-surface patterns have fascinated researchers for a long time, and illuminating qualitative descriptions and explanations go back decades \cite{banerjee1994,longuet-higgins96,brocchini2001a}. 
Turbulence-made patterns on the surface are conventionally divided into three main categories: `dimples' are relatively small circular depressions of the surface where a surface-orthogonal `bathtub' vortex is attached, `scars' are long, narrow surface indentations that appear together with a strong mainly horizontal vortex tube underneath, and `boils', smooth, expanding areas of elevated surface height due to a strong upwelling of water from the deep. 

Recent studies \cite{babiker2023,aarnes2025,babiker2026} have demonstrated that the prevalence of surface features is highly correlated with flow statistics averaged over a suitably large area, demonstrating what the features can tell about \emph{when} flow events happen, but not \emph{where}. In the present study, we answer the latter by introducing a method from spatial statistics which allows us to uncover new physics, in particular a symmetry in the individual roles of dimples and scars.

\subsection{Background and motivation}

When seeking to infer sub-surface flow features from surface observations, some qualitative statements are fairly obvious: clearly the surface imprints can give \emph{some} information about sub-surface turbulence, equally clearly there are aspects which the surface cannot reveal, specifically small scales, that do not have observable surface effects. Moreover, intuition indicates that the further from the surface, the less flow information can be obtained this way, although what the depth dependence of correlations between surface imprints and concomitant flow features might be, is not clear \emph{a priori} and has been a topic of recent study \cite{aarnes2025,babiker2026,Ruth2026}. 

Far less obvious, however, is the question of how flow and surface imprints are connected in time and space. Specifically, dimples and scars have been observed to occur in connection with large upwelling events \cite{longuet-higgins96}, yet especially a dimple can persist for a long time, outlive the upwelling that caused it and travel a considerable distance from where it originated before disappearing. 
If one is to infer sub-surface properties from surface observations, this nonlocality between `cause' and `effect' (in a loose sense, as discussed in \cite{aarnes2025}) must be taken into account. Given an observation of a dimple or a scar, and assuming they originate from an upwelling, when did said upwelling event most likely occur---past, present or even future---and how far away?

The question has practical consequences because large upwelling events are highly effective in bringing fresh water to the surface from below, thus keeping the surface out of thermal and chemical equilibrium with the atmosphere so that heat and gas can be exchanged, a process termed surface renewal \cite{kermani2009}. A measure of the local flux of water to and from the water surface is the \emph{surface divergence}, which is particularly strong in upwelling (positive) and downwelling (negative) regions. For small surface deformations
and continuous interface, the surface divergence is (modulo negligible zero-mean corrections due to surface slope) $\beta_s(x,y,t) = \partial_xu+\partial_yv=-\partial_z w$ evaluated at the surface level, with $(u,v,w)$ the velocity components in $(x,y,z)$ directions with $z$ being vertical upwards. 
(For flows where strong turbulence causes the interface to overturn, form sharp cusps or disconnect, the approximation breaks down; see \cite{brocchini2001dynamics}).
Gas flux across the air-water-surface has frequently been modelled in terms of $\beta_s$ \cite{McKenna2004,banerjee2004,turney2005,janzen2010,turney2013,pinelli2022}, hinting that observing surface features could be a proxy for observation of $\beta_s$, and hence the flux of gas.

Analysing data from direct numerical simulations performed at the University of Minnesota \cite{guoInteractionDeformableFree2010, xuanConservativeSchemeSimulation2019}, \citet{babiker2023} found that the total number of observable dimples on the surface in the computational domain was highly correlated in time with the mean-square surface divergence, $\MSbetas(t)$ (overbar denotes the average over a gridplane on a surface adhering grid), with a slight lag in time between surges in surface divergence and number of dimples, reflecting that surface-attached vortices form a little after the appearance of a major upwelling. Recently, the same observation was made also using experimental data \cite{babiker2026}, where it was demonstrated that scars can be an equally effective proxy albeit with insignificant lag in time, and that the dimples and scars are also strongly correlated with the mean-square of the corresponding divergence field, $\beta(x,y,z,t) = \partial_xu+\partial_yv$, averaged plane-wise \textit{below} the surface.

The close connection between the density of dimples and mean-square surface divergence exemplifies the non-local interrelations at play, highlighted by the observation \cite{khakpour2012} that the surface-attached vortices are sites where scalar gas transfer is particularly \emph{weak}. Hence they occur simultaneously with surges in surface divergence (or a little after), but not in the same \emph{place}. In contrast, parts of near-surface vortices which were horizontal were found to lie next to sites of high surface flux; these are the vortices that create scars above them, and often occur at the periphery of upwelling boils \cite{aarnes2025}. A more local dependence between hotspots of surface divergence and scars is thus hinted at, both in time and in space. 

The analyses of the link between surface features and divergence by Babiker et al.\ \cite{babiker2023,babiker2026} were \emph{global} in nature, in that the entire flow domain or measurement area was considered and features were simply counted. 
Although a strikingly high correlation was found in said study, the clear spatial nonlocality between dimples and upwellings leaves something to be desired. One issue is in terms of generality, in that the numerical domain size of \citet{babiker2023} is rather arbitrary. Would the correlation have been equally strong with a smaller or larger interrogation window? This question is addressed briefly in \cite{babiker2026} (Appendix E), where it was observed that dimples required the entire domain to be an effective proxy for the mean-squared surface divergence, whereas if scars were used as proxy only a smaller portion of the surface area was sufficient.

Such observations motivate the present study, where we aim to answer the following questions: If we know the location of scars and dimples, how far away from them---spatially and temporally---is the concomitant surface divergence hotspot (if there is one) most likely to be? Or conversely: having detected a hotspot in surface divergence, how far from it can we expect the associated scars and dimples to appear? And moreover, are there other fields which have a similar spatial and temporal connection to the dimples and scars as the mean-square surface divergence?

To answer these questions, we introduce a method from spatial statistics novel to the fluid mechanics community, which yields a systematic approach to considering local and global aspects of the surface features and the flow. We model surface features (dimples, scars) as point processes and parametrise the temporal and spatial shift with respect to the surface divergence field. In turn, the parameters are estimated using data from direct numerical simulations (DNS) performed by A.~Xuan and L.~Shen at the University of Minnesota \cite{guoInteractionDeformableFree2010, xuanConservativeSchemeSimulation2019}, to arrive at an optimal model, both at the surface and at increasing distance from below the surface.  
We also isolate the effect of positive and negative surface divergence, bringing new insights in how dimples are distributed near regions of up- and downwelling, respectively.
Furthermore, we perform these modelling steps for the coupling of surface features to a second scalar field, namely the vertical vorticity component, $\omz = \partial_xv-\partial_yu$. When evaluated at the surface $z=0$ the symbol ${\omz}_s$ is used. The consideration adds an edifying symmetry to the analysis vis-à-vis the local and global models for horizontal divergence: Whereas the surface flux is low at dimple location and high at scar location, as detailed above, the vertical vorticity is high where we find dimples---recall that dimples are imprints of vortices that attach perpendicular to the surface---and low at scar location (as documented in \cite{aarnes2025}).

\subsection{Outline}

The paper is organized as follows: Section \ref{sec:datasets} concerns the DNS datasets, six datasets for free-surface turbulent flow, initialized with two different Reynolds numbers and three different Weber numbers. The data is documented in previous publications, and we only include the necessary details here. We introduce the point pattern approach from spatial statistics in Section \ref{sec:Point_Pattern_statistics}. 
We present the statistical method in some detail before describing how we apply it to our specific case. We present our main results in Section \ref{sec:results}, where we also discuss how the local models work for the different surface features and scalar fields, both at and below the surface. Conclusions are drawn in Section~\ref{sec:conclusions}.


\section{Data and processing}
\label{sec:datasets}
The statistical methods we present are applied to six different datasets of turbulent free-surface flow. The datasets have been extensively described and previously used in \citet{aarnes2025, babiker2026, moen2026}, and therefore only a short summary is included here.

The datasets are from Direct Numerical Simulations (DNS) of turbulence generated in a box interacting with a deformable free surface (see \cite{guoInteractionDeformableFree2010, xuanConservativeSchemeSimulation2019, aarnes2025} for information on the simulation setup). The flow in the simulation is agitated by random linear forcing (see \cite{rosalesLinearForcingNumerical2005}), which is gradually dampened as the surface is approached, with a region near the surface experiencing no forcing. The six datasets are initialized with two different Reynolds numbers and three different Weber numbers. 

Flow details are listed in Table \ref{tab:flopProp}, where the datasets are denoted cases 1--6, and all values are given as dimensionless groups or non-dimensional scales. The quantities listed include the Taylor Reynolds number, \(\Rel = u' \Taylor / \nu\), the turbulent versions of the Reynolds number, Froude number and Weber number, $\Reint = 2 \Lint u'/\nu$,  $\Frt = u'/\sqrt{2g \Lint} $ and $\Wet = 2\rho u^{\prime 2} \Lint /\sigma $, respectively, and the integral time scale $\Tint=2\Lint/u'$. Flow parameters for kinematic viscosity, $\nu$, gravitational constant, $g$, density, $\rho$, and surface tension, $\sigma$, are set in simulation initialization, while the relevant length and velocity scales, $\Lint$, $\Taylor$ and $u'$ are computed \emph{a posteriori}. The latter are the integral length scale, $\Lint = \Taylor \Rel/30$, Taylor length scale, \(\Taylor = u' \sqrt{15 \nu/ \epsilon}\), and representative velocity $u' = \sqrt{\frac13\mean{u_i u_i}}$, where $\epsilon = 2 \nu \mean{s_{ij}s_{ij}}$ is the dissipation with $s_{ij} = (\partial_i u_j +\partial_j u_i)/2$, $\langle\cdots\rangle$ denotes average taken in time and an overbar denotes spatial average taken over a horizontal gridplane. Finally, we also include the Kolmogorov length scale $ L_K = \left(\nu^3/\epsilon\right)^{1/4} $ for completeness. The quantities are calculated at a depth corresponding to bulk flow.
Length and velocity scales in Table \ref{tab:flopProp} are normalized with the characteristic length and velocity of the simulation \cite[details in][]{aarnes2025}. 
All cases lie within ``Region 0" of the classification by \citet{brocchini2001a}, characterized as “weak turbulence” with only very small surface deformations \citep[see ][ for more details]{babiker2026}. (The values in Table \ref{tab:flopProp} differ slightly from those reported in \cite{babiker2026} due to non-dimensionalization and a different estimation procedure for $\Lint$. This has no bearing on any of our analysis or conclusions.)

\begin{table}
    \centering
    \begin{tabular}{p{2.5em} p{2.5em} p{2.5em} p{2.8em} p{2.5em} p{2.5em} p{2.8em} p{2.5em} p{2.5em} p{3.5em} }
    \hline
        Case & $\Rel$ & $\Reint$& $\Frt$ & $\Wet$   & $\Tint$   & $u'$  & $\Lint$ & $\Taylor$ & $L_K$  \\
        \hline
        1 & 81 & 433 & 0.012 & $\infty$ & 9.4 & 0.136 & 0.64 & 0.24  & 0.0135 \\
        2 & 74 & 368 & 0.012 & 0.38 & 8.8 & 0.130 & 0.57 & 0.23 & 0.0135 \\
        3 & 82 & 452 & 0.013 & 0.26 & 8.75 & 0.144 & 0.63 & 0.23 & 0.013 \\
        4 & 44 & 131 & 0.012 & $\infty$ & 8.85 & 0.122 & 0.54 & 0.36 & 0.028 \\
        5 & 50 & 164 & 0.012 & 0.44 & 9.1 & 0.134 & 0.61 & 0.37 & 0.027  \\
        6 & 47 & 150 & 0.012 & 0.19 & 9.2 & 0.128 & 0.59 & 0.37 & 0.027  \\
        \hline
    \end{tabular}
    \caption{Turbulent properties for the DNS cases. From left: case name, Taylor Reynolds number, turbulent Reynolds number, turbulent Froude number, turbulent Weber number, integral time scale, representative velocity, integral length scale and Kolmogorov length scale. Length and velocity scales normalised with the characteristic length and velocity from the simulation initialization. 
    }
    \label{tab:flopProp}
\end{table}

To detect surface features in the data and classify them as either dimples or scars, we use direct detection from the full flow field for the dimples and indirect identification from surface height data for the scars. The direct detection of dimples is done by identifying the vortical structures in the flow according to the $\lambda_2$ criterion by \citet{jeongIdentificationVortex1995} (discussed in detail in \cite{aarnes2025}), and identifying the imprints of surface-attached vortical tubes as dimples. Strictly speaking, identifying surface-attached vortices is not equivalent to detecting their surface indentations: the direct approach precludes false-positive dimple detections, the effect of which is discussed in Appendix~\ref{app:detected_dimps}.
As we do not have an equivalent detection of scars from the flow field, we use rather the wavelet-based detection algorithm documented in \Babiker, originally developed only for dimple detection and later extended to also include scars \cite{aarnes2025}. The method uses the so-called Mexican-hat wavelet on the surface elevation data and sorts the resulting coherent structures by eccentricity, classifying high-eccentricity structures as scars. The method is accurate for detecting surface features, particularly from high temporal resolution DNS data, and it has also been applied successfully to experimental data with much lower temporal resolution \cite{babiker2026}. Note, however, that some false positives in scar detection should be expected, in particular for small structures where some dimples are misclassified as scars (as discussed in Section 5.6 of \cite{aarnes2025}).


\section{Methods: Point pattern model and inference}
\label{sec:Point_Pattern_statistics}
In this section, we introduce point pattern models from spatial statistics, which we use to develop and estimate local models for the association between dimples, scars, surface divergence and vorticity. For our purposes, the method consists of three steps: (1) We take a point pattern analysis approach, that is, we model surface features as points or locations at the surface, (2) we parametrize the temporal and spatial shift (i.e., the local support) between locations and field intensities and (3) we use the DNS datasets to estimate relevant model parameters. The flow we consider is a free-surface turbulent flow, where the turbulence is strong enough to disturb the surface to the extent that dimples, scars and upwelling boils are visually dominant patterns at the surface. Details of the datasets, including turbulence levels and surface feature detection, are given in Section \ref{sec:datasets}.

\subsection{Point pattern approach}

We start by considering the areas of the surface which we have identified as surface features of interest and classified as either dimples or scars. These features are represented as point observations, prescribed to a single point of reference for each feature. For a dimple we use the centroid of its area; for a scar, which is elongated and often crescent-shaped, we use the midpoint of its centerline identified with a skeletonisation method \citep[][section 5.4]{aarnes2025} 
(a statistical theory that treats scars as line segments is beyond the scope of the present work).
The resulting surface feature data is a \emph{point pattern}, that is, a collection of locations in the observation window at each time step categorized as either a dimple or a scar.
Point pattern analysis has an extensive literature (see \cite{Illian2008StatisticalAA} for an introduction). 
For each point in time the point pattern is described by the list of coordinates $N(t)=\{(x_{1,t},y_{1,t}),...,(x_{|N(t)|,t},y_{|N(t)|,t})\}$ of all $|N(t)|$ points in the observation window $W$.

Our interest is to relate surface features represented as point patterns to flow fields. We use surface divergence and vertical vorticity as scalar flow fields to which we relate the surface features, denoting a scalar field $\phi$ (representing either $\omz$ or $\beta$). It is evaluated at location $(x,y)$ and time $t$ in a plane at some depth co-ordinate $z$, as $\sfield(x,y,t;z)$. We suppress the coordinate $z$ in the following so as not to confuse it with model parameters, and let it be understood that $\phi$ can be evaluated at the surface (subscript `s') or at some constant depth $z<0$.
To connect $\sfield(x,y,t)$ to the point pattern, we use an inhomogeneous Poisson point pattern model, which is characterized by an intensity function $\intensity(x,y,t)$ (see Chap.~3.4 and 6.4 in \cite{Illian2008StatisticalAA}). Given the intensity function, a point pattern $N$ follows an inhomogeneous Poisson point process if the following conditions are satisfied,
\begin{itemize}
    \item For any sub-area $\subWindow\subset\obsWindow$, the number of points in $\subWindow$, $|N(t,\subWindow)|$, is Poisson distributed, with $E[|N(t,\subWindow)|]=\int_\subWindow \intensity(x,y,t)dxdy$, where $E[\cdot]$ denotes the expected value.
    \item For any disjoint sub-windows, $\subWindow_1\subset\obsWindow$ and $\subWindow_2\subset\obsWindow$, the number of points in the sub-windows is independent, i.e. the random variables $|N(t,\subWindow_1)|$ and $|N(t,\subWindow_2)|$ are independent.
\end{itemize}
In particular, this means that $|N(t,\subWindow)|$ has distribution
\begin{equation}
    P(|N(t,\subWindow)|=m)=\frac{e^{-\int_A\nu(x,y,t)dA}\left(\int_A\nu(x,y,t)dA\right)^m}{m!}.
\end{equation} 
Note that this is only a valid probability if $\int_A\nu(x,y,t)dA\geq0$, implying that any valid intensity function $\nu(x,y,t)\geq0$ everywhere. We aim to model the intensity of dimples and scars separately, denoting the respective intensities as $\intensity\subd$ and $\intensity\subs$. When it is unspecified which surface feature we are studying, we drop the subscript. The inhomogeneous Poisson point pattern model is the ``standard" model to use in situations when studying the relationship between a point pattern and a scalar field and the natural choice for our purposes, arguably being the simplest, most transparent and most documented alternative.
For some diagnostics on the model assumptions, see Appendix \ref{app:diagnostics}.

The first bulletpoint above means that $\intensity$ can be understood as an expected number density distribution for points.
The intensity $\intensity$ has unit `per unit area', but all quantities are rescaled by a reference length to be dimensionless \cite{aarnes2025}; the DNS data we use has dimensions $2\pi\times2\pi$ in these units, so a spatially homogeneous distribution with expectation $|N(t)|$ total points would correspond to $\nu(t)=|N(t)|/(4\pi^2)$.

\begin{figure}
    \centering
    \begin{overpic}[width=1.0\linewidth]{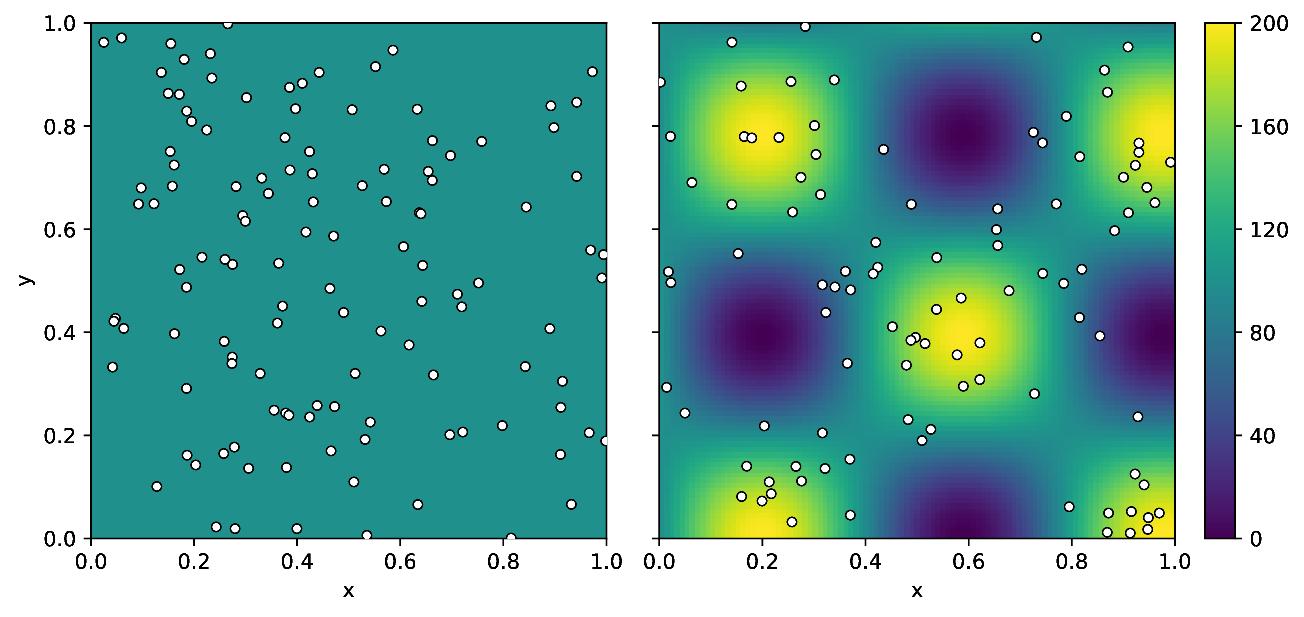}
        \put (9, 42) {\color{white}\textbf{(a)}}
        \put (53, 42) {\color{white}\textbf{(b)}}
    \end{overpic}
    \caption{Two example realisations of point patterns on the unit square with positions distributed randomly according to the full-window model with $\intensity=100$ (a) and the inhomogeneous Poisson model with $\intensity(x,y) = 100[\sin(8x)\cos(8y)+1]$ (b).
    }
    \label{fig:poisson_point_pattern_example}
\end{figure}

To illustrate the role of the intensity function in an inhomogeneous Poisson point process, we provide two examples, each simulated on a window W with unit side lengths: the first is a constant intensity $\intensity = 100$ (i.e.\ a homogenous Poisson process), the second is a spatially varying intensity which we specify as $\intensity(x,y) = 100[\sin(8x)\cos(8y)+1]$. Figure \ref{fig:poisson_point_pattern_example} depicts the intensity functions, each plotted together with a corresponding simulated point pattern; visual inspection is consistent with the interpretation of $\intensity(x,y)$ as an expected point density.  

\subsection{Models for local flow properties and point pattern intensity}

With the concepts of flow field, $\sfield(x,y,t)$, and inhomogeneous Poisson point process following intensity function, $\intensity(x,y,t)$, in place, we may introduce the remaining string of concepts for our local model. These are ``local flow properties",  ``local support", and---specifically tied to our application---the ``local variance". 

With \textit{local flow properties}, we mean properties of the flow field which we arrive at by a transformation restricted to a spatially local scale. The local flow property is parametrized with this local scale, which we denote the \textit{local support}, $r$.
An example of a transformation from a global to a local scale is to go from considering the full-window (i.e., global) mean of a flow field
\begin{equation}
    \overline{\sfield}(t) = \frac{1}{A_W}\int_{W}\sfield(\xi,\eta,t)d\xi d\eta \, ,
\end{equation}
where $A_W$ is the area of the observation window $W$, to consider the local mean,
\begin{equation}
    \label{eq:local_mean}
    \overline{\sfield}_r(x,y ,t) = \frac{1}{\pi r^2}\int_{B_r(x,y)}\sfield(\xi,\eta,t)d\xi d\eta,
\end{equation}
where $B_r(x,y)$ is the disk with radius $r$ around $(x,y)$. The resulting local flow property, the local mean $\overline{\sfield}_r$ at location $(x,y)$, is a function of $\sfield$ only within the area $B_r(x,y)$.
It is apparent that the local mean $\overline{\sfield}_r(x,y ,t)$ retains more information than the global mean $\overline{\sfield}(t)$, since the latter is an average over the entire observation window, and hence, contains no local information. 

\Babiker{} found a strong correlation between the number of dimples in the whole observation window and the surface divergence squared, averaged over the entire observation window. The correlation was maximal when there was a lag $\tau$ between $\MSbetas$ and the dimple count. In our framework this suggests a simple model for dimples, namely $\nu\subd(t)\propto\MSbetas(t+\tau)$, i.e., $\intensity$ is independent of the location $(x,y)$, resulting in a \emph{global} model. 
Going forward, when exploring different local models we allow a time shift $\tau$, but do not specify whether it is positive or negative (i.e., we do not specify if the surface features lead or lag the relevant flow field).
The more general form chosen for our global models is 
\begin{equation} \label{eq:global_model}
\nu(x,y,t;a,b,\tau)=a\bigl[\MSphi(t+\tau)\bigr]^b
\end{equation}
with three parameters $a,b$ and $\tau$.

To facilitate the development of a local model we define the \emph{local variance} of $\sfield$ as the empirical variance of the flow field $\sfield$ in a circle of radius $r$ around $(x,y)$, 
\begin{equation}
    \label{eq:sigphi}
    \sigfi(x,y,t;r,\tau) = \frac1{\pi r^2}\int_{B_r(x,y)}\left(\sfield(\xi,\eta,t+\tau)-\overline{\sfield}_r(x,y,t+\tau)\right)^2d\xi d\eta
\end{equation}
where $\overline{\sfield}_r$ is the local mean, defined in Eq. \eqref{eq:local_mean}.
When $W$ covers the full computational domain, the imposed conservation of mass and angular momentum together with the periodic boundary conditions imply that $\overline{\beta}$ and $\overline{\omega}$ are zero, hence
\begin{equation}
    \sigfi \buildrel{B_r\to W}\over{\longrightarrow} \MSphi,
\end{equation}
and $\sigfi$ is a local generalisation of $\MSphi$, noting that in the smaller domain $B_r$, $\overline{\sfield}_r$ is not in general negligibly small.
In practice, our data are discrete, so we use the empirical variance for the discretisation points within the radius $r$ as the local variance.

For the surface divergence, it is also instructive to consider the contributions from the positive and negative parts of the surface divergence separately, as these can be tied to regions of upwelling (positive $\beta$) and downwelling (negative $\beta$) \citep[e.g.,][]{guoInteractionDeformableFree2010}. We split $\beta(x,y,t)$ into two fields according to its sign, each defined piecewise according to
\begin{equation}
    \label{eq:pos_neg_divergence_def}
    \beta^{\pm}(x,y,t)=\max\{\pm\beta(x,y,t),0\}.
\end{equation}
Here, the local flow properties of interest are the local means $\overline{\beta^\pm}_r$, defined as in Eq.~\eqref{eq:local_mean}.

\begin{figure}
    \centering
    \begin{overpic}[width=0.8\textwidth, grid=off, tics=5]{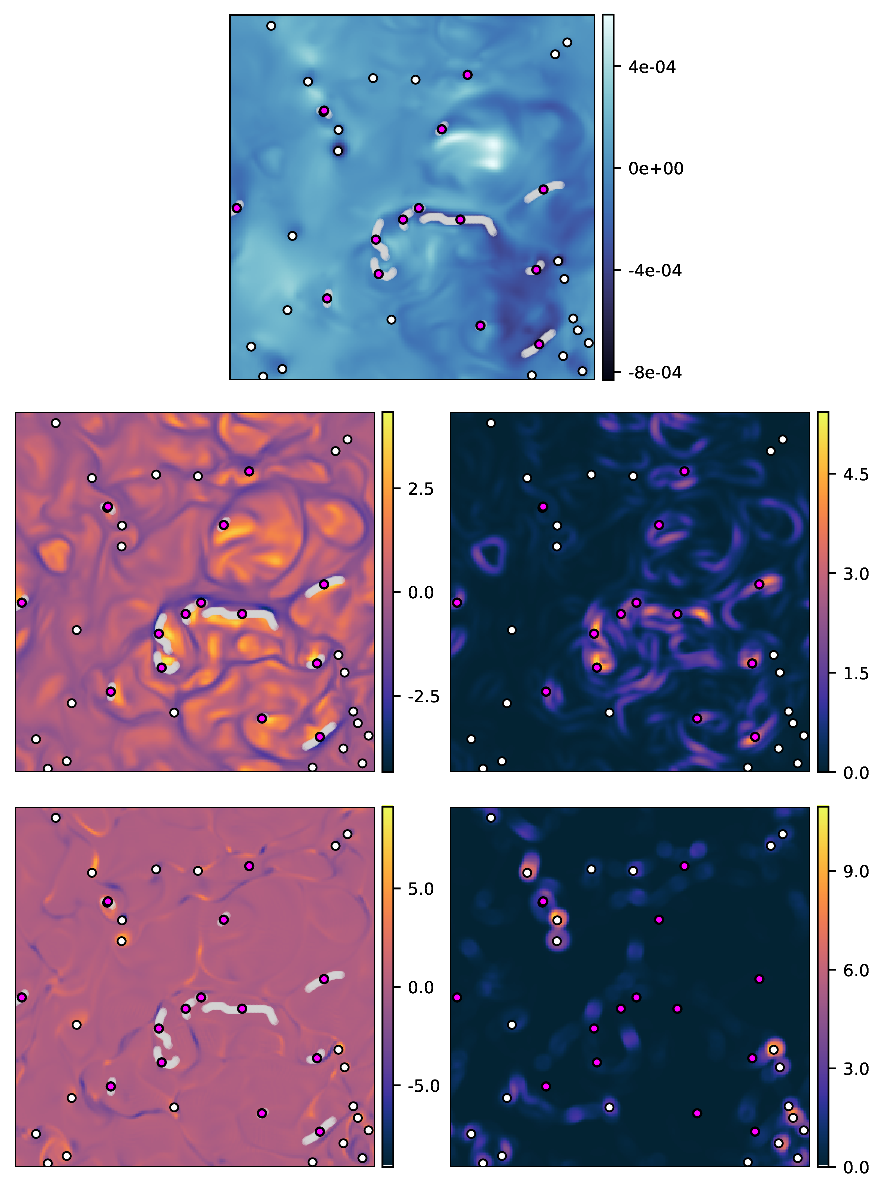}
            \put (22,94.5) {\color{white}\textbf{(a)}}
            \put (4,61) {\color{white}\textbf{(b)}}
            \put (40.5,61) {\color{white}\textbf{(c)}}
            \put (4,28) {\color{white}\textbf{(d)}}
            \put (40.5,28) {\color{white}\textbf{(e)}}
    \end{overpic}
    \caption{Qualitative illustration of our procedure, at an arbitrarily chosen timestep. The displayed fields are (a) surface elevation $\eta(x,y)$, (b) surface divergence $\beta_s(x,y)$, (c) local variance $\sigbeta(x,y)$ of $\beta_s$, (d) vertical vorticity $\omz_s(x,y)$ and (e) local variance $\sigom(x,y)$ of $\omz_s(x,y)$. Overlaid are dimple centers (white dots), scar centers (pink dots) and scar centerlines (grey lines). The local variance fields are computed with local support $r=0.23$ and time shift $\tau=0$. Each field is plotted for the full domain size.}
    \label{fig:surface_example}
\end{figure}

Figure \ref{fig:surface_example} illustrates local variance for an example snapshot from the simulation data. All panels are plots at the free-surface level spanning the computational domain with the point pattern due to the dimples and scars identified at the surface (panel a) overlaid as white and pink circles, respectively. The flow fields $\beta_s(x,y)$ and $\omz_s(x,y)$ are plotted in panels (b) and (d) and the corresponding local variance of the same respective fields, $\sigbeta(x,y;r)$ and $\sigom(x,y;r)$ with $r=0.23$ (about $4\%$ of the side length of the domain) and no time shift. 
Visual inspection of (c) and (e) hint at a close local relation between $\sigbeta$ and scar locations, and similarly between $\sigom$ and dimple sites.

We can construct a local model of the Poisson point process intensity as a function of the flow fields.
The standard approach to modelling intensity, as described in \cite{GLM_fahrmeir} and \cite{GLM_mcCullagh}, is to use a log-linear intensity regression model,
\begin{equation}
    \label{eq:log_linear_model}
    \log \intensity(x,y,t)=b_0+b_1 \zeta(x,y,t;r,\tau),
\end{equation}
where $\zeta$ is some local flow property, and $b_0$ and $b_1$ are real constants. This approach is practical, as it ensures non-negative intensity, is conducive to numerical stability during inference, and allows us to use standard statistical software such as the R-package \texttt{spatstat} \cite{spatstat}. The local variance is non-negative by definition (and strictly positive in our data for all cases), hence we may use $\zeta=\log{\sigma^2_\sfield}$, which gives a direct generalisation of Eq.~\eqref{eq:global_model},
\begin{equation}
\label{eq:power_law_model}\intensity(x,y,t;a,b,r,\tau)=a\bigl[\sigma^2_{\sfield}(x,y,t;r,\tau)\bigr]^{b},
\end{equation}
where $a=\exp b_0$ is required to be strictly positive, and $b=b_1$. 
We refer to Eq.~\eqref{eq:log_linear_model} as the former as the log-linear model and \eqref{eq:power_law_model} as the power law model. In the case where we study the local mean of positive and negative surface divergence separately, our domain contains large areas where $\beta^\pm=0$, in which cases we therefore use the log-linear intensity $\zeta=\overline{\beta^\pm}_r$ directly. 
Note that a local model means there will be a higher intensity of scars or dimples in an area where $\zeta$ is large if $b_1$ is positive, while if $b_1$ is negative a local model implies the intensity is reduced in areas where $\zeta$ is large.

Two surface feature types (dimples, scars) and two fields ($\beta$, $\omz$) gives four models to analyse. With the addition of the models for separate contributions from positive and negative surface divergence, this yields a total of six statistical models.
The models are summarised
in Table \ref{tab:Models}. It is understood that each model comes in local and global form, and that parameters $a,b$ (or $b_0,b_1$ for $\beta^\pm$) will take different values for each model. In Section \ref{sec:results} we estimate the local support $r$ and the time shift $\tau$ for each of these models, and compare their predictive performance to their global counterparts. All models are analysed at the surface $\sfield_s(x,y,t; ...)$ and at depths $z<0$ $\sfield(x,y,t;z,...)$. For a discussion of other local flow properties, see Appendix \ref{app:model_selection}.

\begin{table}[]
    \centering
    \caption{Overview of local models.}
    \begin{tabular}{cccc}
     \hline
     Model name & Surface feature & Flow field & Intensity $\intensity$\\
     \hline
     DimpDiv & Dimples & Surface divergence & $\intensity\subd=a(\sigma^2_\beta)^{b}$\\
     DimpVort & Dimples & Vertical vorticity & $\intensity\subd=a(\sigma^2_{\omz})^{b}$\\
     ScarDiv & Scars & Surface divergence & $\intensity\subs=a(\sigma^2_\beta)^{b}$\\
     ScarVort & Scars & Vertical vorticity & $\intensity\subs=a(\sigma^2_{\omz})^{b}$\\
     DimpDivUp & Dimples & Positive surface divergence & $\log\intensity\subd=b_0+b_1\overline{\beta^+}_r$\\
     DimpDivDown & Dimples & Negative surface divergence & $\log\intensity\subd=b_0+b_1\overline{\beta^-}_r$\\
     \hline
     \end{tabular}
    \label{tab:Models}
\end{table}

\subsection{Statistical inference}

We estimate the parameters $a,b,\tau$ and (for local models) $r$ by Maximum Likelihood (ML) estimation. We will denote the parameters collectively by $\theta$, and we emphasize that these are now the variable quantities, while the surface features $N$ are fixed. ML estimation seeks to maximize the likelihood function, $L(\theta;N)=f(N;\theta)$, the probability density of observing the surface features $N$ with parameters $\theta$. 
Maximising $L(\theta;N)$ produces the parameters $\theta$ that make the given observation $N$ most probable.
Under an inhomogeneous Poisson point pattern model the likelihood function for a single snapshot at time $t$ is given by
\begin{equation*}
    L\bigl[\theta;N(t)=\{(x_{1,t},y_{1,t}),...,(x_{|N(t)|,t},y_{|N(t)|,t})\}\bigr]=\exp\left(-\int_W\intensity(x,y,t;\theta)dA\right)\prod_{i=1}^{|N(t)|}\intensity(x_{i,t},y_{i,t},t;\theta),
\end{equation*}
as described in Chap.~3.4.1 of \cite{Illian2008StatisticalAA}, where $\intensity$ is the intensity of the pattern, either following a global model (Eq. \eqref{eq:global_model}) or a local model (Eq. \eqref{eq:power_law_model}). In practice our data set is discrete and consists of $n$ independent snapshots $\{N(t_1),...,N(t_n)\}$ with corresponding likelihood
\begin{align*}
    L\bigl[\theta;\{N(t_1),...,N(t_n)\}\bigr] &=
    \prod_{j=1}^nL\bigl[\theta;N(t_j)=\{(x_{1,t_j},y_{1,t_j}),...,(x_{|N(t_j)|,t_j},y_{|N(t_j)|,t_j})\}\bigr] \\&=
    \exp\left(-\sum_{j=1}^n\int_W\intensity(x,y,t_j;\theta)dA\right)\prod_{j=1}^n\prod_{i=1}^{|N(t_j)|}\intensity(x_{i,t_j},y_{i,t_j}, t_j;\theta).
\end{align*}
We denote the estimated parameters by adding a hat, $\hat\theta$.

We make use of two sources of data for each DNS case: The surface features in the observation window $\obsWindow$ at snapshots for times $\{t_1,...,t_n\}$ and the relevant flow field in $\obsWindow$ at corresponding snapshots for time intervals $\{(t_1+\lagmin,t_1+\lagmax),...,(t_n+\lagmin, t_n+\lagmax)\}$. 
A measure of the time the turbulence remains correlated with itself (`turbulent memory') is the integral timescale $T_\infty$, introduced 
in the previous section. Snapshots separated in time by at least $T_\infty$ can be considered approximately independent. 
Snapshots $\{t_1,...,t_n\}$ are set $t_1=-\tau_{\min}$, $t_k=t_1+(k-1)T_\infty$, and the number of snapshots is limited by $t_n+\tau_{\max}<T$, where $T$ is the total length of the time series. 
The timeshift limits $\lagmin$ and $\lagmax$ must be set large enough that 
the ML estimate is contained in the interval $\lagmin\leq\tau\leq\lagmax$, without severely limiting $n$.
We determine $\tau_{\min}$ and $\tau_{\max}$ by trial and error. For details on snapshot counts, diagnostics on independence, and discussion of sensitivity to the use of $T_\infty$ as time step, see Appendix \ref{app:snapshots_independence}.

For $r$ we allow the full available range up to half the domain size, approximately $0\leq r\leq 5L_\infty$ for all DNS datasets, before the circle of radius $r$ overlaps itself due to the periodic boundary conditions. For the maximum value of $r$ the local and global models are practically the same. 

We maximize the likelihood function only to estimate the parameters. The numerical values of the likelihood function are not interesting in an absolute sense, but the shape of the likelihood is informative and likelihood differences are commonly used to compare the performance of different models (often in the modified form of an information criterion like the AIC, see \cite{GLM_fahrmeir}).
To gain understanding of how the local model performs with sub-optimal choices of $r$ and $\tau$, we use the profile likelihood, $L_p(r,\tau;N)=\max_{a,b}L(r,\tau,a,b;N)$, meaning that for each $r,\tau$ we use the values of $a$ and $b$ which gives maximum likelihood. This choice of metric enables us to make a heatmap showing how the likelihood varies for different choices of $r$ and $\tau$.
We use the profile likelihood function to compare the local and global models.

\section{Results}
\label{sec:results}

In this section, we present the results for optimal parameter values estimated for the models from spatial statistics. Our main focus is on coupling the surface features (dimples and scars) to the divergence and vorticity fields at the surface which we consider in Sec.~\ref{sec:results_surface}, but we also include results that suggest the utility of local models for predicting the corresponding sub-surface fields from the detected dimples and scars, in Sec.~\ref{sec:results_sub-surface}.

\subsection{Model performance: surface fields}
\label{sec:results_surface}

\begin{figure}
    \centering
    \begin{overpic}[width=\textwidth]{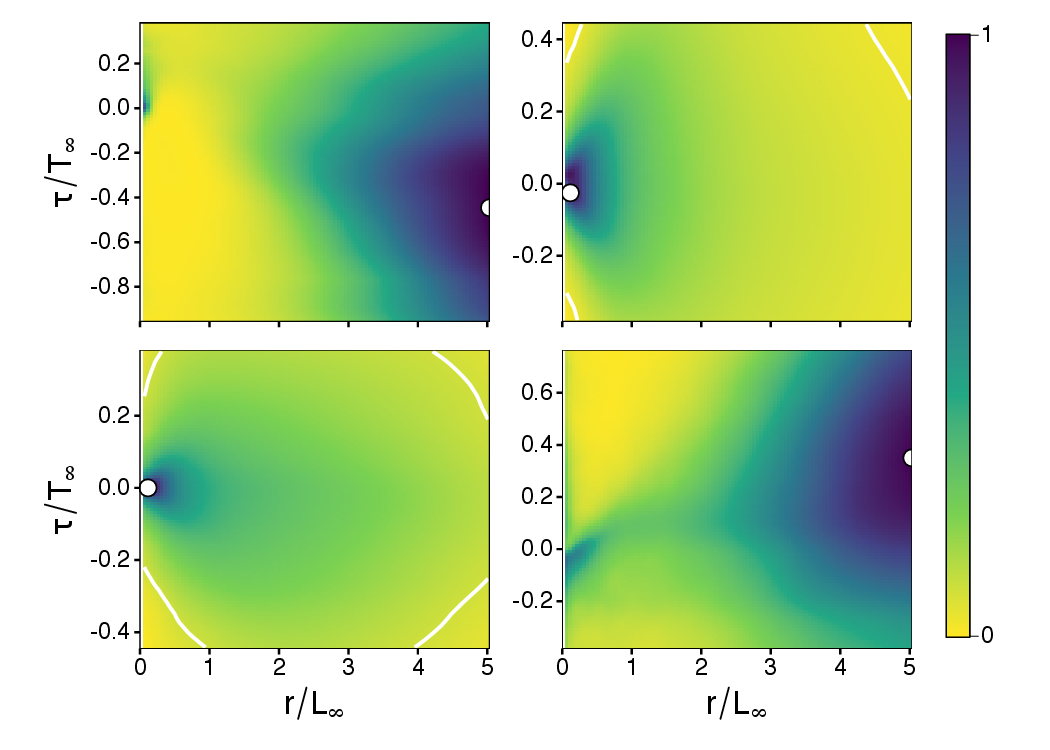}
            \put (15,65) {(a)}
            \put (57.5,65) {(b)}
            \put (15,32) {(c)}
            \put (57.5,32) {(d)}
    \end{overpic}
    \caption{Normalized profile likelihood for the models (a) DimpDiv, (b) DimpVort, (c) ScarDiv and (d) ScarVort, for Case 1 (see Table \ref{tab:flopProp}). Larger values indicate better performance. Spatial support $r$ and temporal shift $\tau$ are normalized with the integral length and time scales of the turbulence, respectively. 
    The white contour indicates where performance equals that of a global model.
    }
    \label{fig:likelihood_surface}
\end{figure}

\subsubsection{Models for local variance of divergence and vorticity}
\label{sec:results_surface_variance}

Figure \ref{fig:likelihood_surface} displays the estimated profile likelihood for the four models relating dimples and scars to the local variance surface divergence and vertical vorticity for Case 1 (case details in Table \ref{tab:flopProp}; model details in Table \ref{tab:Models}; the same figures for Cases 2-6 can be found in the supplementary materials). Panels (a) and (b) couple the dimples to the surface divergence and vertical vorticity field, respectively, while figures (c) and (d) do the same for scars. As mentioned in Section \ref{sec:Point_Pattern_statistics}, the absolute numerical values of the profile likelihood hold little information in themselves and cannot be directly compared from case to case. We therefore normalize the values to lie between 0 and 1, with higher values (closer to one) indicting the best performance. A white dot in each frame marks the estimated scale of local support $r$ and temporal shift $\tau$ for each local model. The white contours in frames (b) and (c) indicate the region where the local model outperforms the global, full-window model. For a comparison of the likelihood values across models, see Appendix \ref{app:model_selection}.

The profile likelihood results suggest that the models can be split into two groups: (1) DimpVort and ScarDiv and (2) DimpDiv and ScarVort. The first group contains local models which clearly outperform the corresponding global models, seen by the fact that the estimated scale of local support $\hat{r}$ is much smaller than the observation window in figures \ref{fig:likelihood_surface}(b) and (c). These models also have estimated time shift $\hat{\tau}$ near zero. Hence, the ``best" local model is localized within a short radius and is approximately instantaneous in time. We may interpret these profile likelihoods to indicate that changes in vertical vorticity occur in immediate spatial and temporal proximity of dimples, while changes in surface divergence occur in immediate spatial and temporal proximity of scars.

The second group, on the other hand, has the global, full-window models outperforming the local models. In the estimated scale of local support displayed in figures \ref{fig:likelihood_surface}(a) and (d), we see that the profile likelihood increases as $r$ grows to its largest admissible value, $r=5\Lint$, yielding a boundary result.
Moreover, the profile likelihood displays a trend of increasing further with $r$ in said figures, indicating that the ``true" $r$ is greater than what can be covered with the observation window in the present simulation. The estimated time shift, for both the ``local" model with $\hat{r}=5\Lint$ and the global model (the two are virtually indistinguishable), is positive for ScarVort, and negative for DimpDiv. For ScarVort, this means that we first see scars appearing \textit{somewhere} at the surface in our observation window, and then a change happens in the vorticity a little later. For DimpDiv, this implies that there is first a change in the surface divergence, and then dimples appear in the observation window a little later.

The estimated likelihood results of the four models in Fig.~\ref{fig:likelihood_surface} can be tied to the physics of the turbulent free-surface flow. The local support with no lag shown by the DimpVort model indicates that surface dimples coincide with regions of strong vertical vorticity, and a small time shift of either sign does not matter much. In itself, this result is not surprising, as the dimples are well known to be long-lived surface manifestation of vortices in the flow that have attached themselves perpendicular to the surface and move slowly, and hence, have strong vertical vorticity \citep{banerjee1994,longuet-higgins96,brocchini2001a}. 
Consistently, laboratory measurements on a quasi-flat surface show that regions of strong surface vorticity persist for times comparable to the integral time scale \citep{qi2025small}. The result is nevertheless welcome as it helps establish confidence in our modelling approach. 

More interesting, however, is the fact that the global full-window model outperformed the local model for the coupling of dimples and surface divergence, suggesting that dimples do not generally live in immediate proximity of regions with large changes in surface divergence, i.e.\ areas of upwelling boils.
After manifesting towards the end of, and at the outskirts of, upwelling events, the long-lived dimples are expelled from the upwelling region by the diverging flow from the boil  and drift towards the flow convergence zones created by downwellings. As the DimpDiv model does not distinguish between effects of positive and negative divergence, the contrasting effects of repulsion and attraction between up- and downwellings are not revealed by the DimpDiv model, resulting in the non-local conclusion illustrated in Fig.~\ref{fig:likelihood_surface}(a).
We return to this question towards the end of this section.

The model ScarDiv shows very similar behaviour as the DimpVort model, that is, no time lag and strong local support for small $r$. The results corroborate earlier observations that scars are adjacent to upwellings. Compared to DimpVort the decline away from zero radius is slower, corresponding to scars and upwellings not quite coinciding and upwellings varying greatly in size.
Moreover, the results suggest that the scars are concurrent with the upwelling and downwelling events, not only at inception but throughout the scars' lifetime: The model rapidly worsens when time lag is not zero, showing that scars, unlike dimples, appear and disappear together with strong surface divergence events. 
Scars are imprints of the horizontal, `smoke-ring'-like vortex that encircles the upwelling \citep{aarnes2025} whose strength immediately varies with the speed of fluid welling up through it. 
Lying between the expanding upwelling boil and the surrounding compressive downwelling region, scars thus mark boundaries between diverging and converging flow zones. 
As these drive inter-scale energy transfer to larger and smaller scales, respectively \citep{wu2026localised}, scars may indicate sites where cascades in both directions occur in close proximity.

The lack of local support for the model ScarVort, which links scars to vertical vorticity, is in line with the observation of \Aarnes\ that scars are located directly above horizontal sub-surface vortices where the vertical vorticity component is very weak. The optimum value of $r$ and $\tau$ for ScarVort corresponds to a global model with positive time shift, which is readily explained in light of the position of scars with respect to dimples. Dimples are surface manifestations of strong vorticity, and scars appear and disappear with the upwelling boils which precede the dimples. Hence, the two should be strongly correlated in time, now with a positive time lag, but only weakly correlated in space. 

Although ScarVort has an optimum $r$ so large as to be effectively global, Figure \ref{fig:likelihood_surface}(d) also shows an additional local maximum in the vicinity of $r/\Lint=\tau/\Tint=0$. The likely causes are a combination of misclassification (a small percentage of dimples erroneously identified as scars) and scars which so closely overlap with dimples that the features are indistinguishable. Both causes lead to a certain ``leakage'' from the results in (b) to (d). 

\subsubsection{Separate models for up- and downwelling}
\label{sec:DivUpDivDown}

\begin{figure}
    \centering
    \begin{overpic}[width=\textwidth]{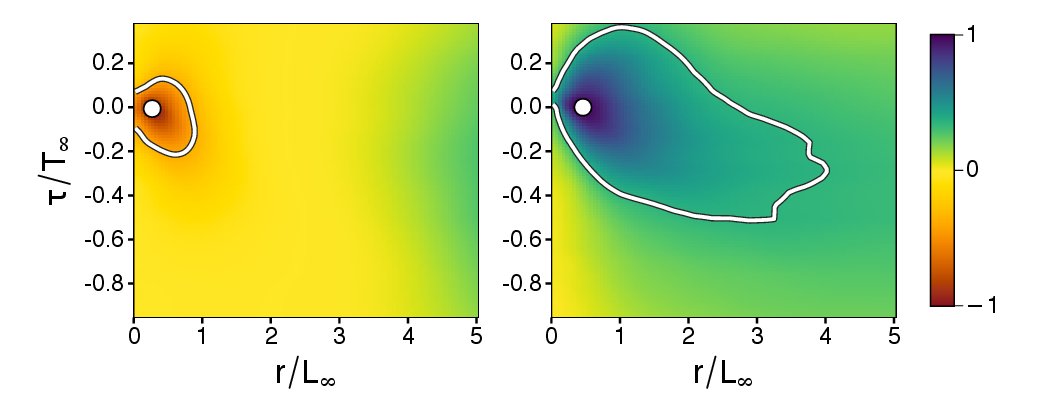}
            \put (15,12) {(a)}
            \put (55,12) {(b)}
    \end{overpic}
    \caption{
    Same as figure \ref{fig:likelihood_surface}, for models (a) DimpDivUp (dimples vs $\overline{\beta^+}_r$) and (b) DimpDivDown (dimples vs $\overline{\beta^-}_r$), for Case 1 (see Table \ref{tab:flopProp}). Likelihoods have been multiplied by the sign of the coefficient $b_1(r,\tau)$.
    }
    \label{fig:signed_analysis_likSurf}
\end{figure}

Figure~\ref{fig:signed_analysis_likSurf} shows the estimated profile likelihood for the two models where we separate the contribution from upwellings (DimpDivUp) and downwellings (DimpDivDown), again for Case 1. To highlight the differences in behaviour compared to DimpDiv, we have multiplied likelihoods by the sign of $b_1(r,\tau)$.
The sign of $b_1$ is essential because a positive value (true of all models in Fig.~\ref{fig:likelihood_surface} throughout the $(r,\tau)$ plane) causes a higher likelihood combined with $b_1<0$ indicates that the expected density of dimples is \emph{lower} than the mean, and areas with high field values predict not where the dimples most likely are, but where they are \emph{not}. 
Figure \ref{fig:signed_analysis_likSurf} then shows how upwellings (high $\beta^+$) are regions where the expected density of dimples is low, and the downwellings are regions where the expected density of dimples is high.

Figure~\ref{fig:signed_analysis_likSurf}(a) thus has a different, in a sense opposite, interpretation from the other panels in figures \ref{fig:likelihood_surface} and \ref{fig:signed_analysis_likSurf}: that is, a local DimpDivUp with small support radius effectively predicts areas \emph{without} dimples.

The interpretation of the opposite trends in the two heatmaps in Fig.~\ref{fig:signed_analysis_likSurf} is clear when considering that up- and downwellings are zones of divergence and convergence of the surface flow. Vortices are transported with the surrounding flow according to Helmholz' theorems, so areas of high positive $\beta$ will repel dimples, whereas $\beta <0$ indicates a convergence zone which attracts them. 
However, for larger $r$, the value of $b_1$ of DimpDivUp becomes positive, showing that the presence of upwelling, for example inside an upwelling boil, is a point in time when the dimple density is increased globally, but a position where it is decreased locally.
The contrast with the conclusions drawn for the DimpDiv model (Fig.~\ref{fig:likelihood_surface}) is also readily explained: A local surge in $\beta^2$ could be either an upwelling or a downwelling making it inconclusive whether the local expected density is increased or decreased. Thus $\beta^2$ is a good indicator of the prevalence of dimples \emph{globally}, but information about where the dimples are more likely to be is lost.

The DimpDiv model analysed in Figure \ref{fig:likelihood_surface}(a) was found to have its optimum for a significant negative time shift. This corresponds to the observations of \Babiker{} that dimples systematically manifest some time after a surge in the surface divergence in the form of upwelling boils. The same lag can be seen to the far right of Fig.~\ref{fig:signed_analysis_likSurf}(a), whereas the strong repulsion of dimples by upwellings---the extremum near $r/\Lint=\tau/\Tint=0$---is local and instantaneous. 

\subsection{Model performance: Sub-surface fields}
\label{sec:results_sub-surface}

\begin{figure}
    \centering
    \begin{overpic}[width=\textwidth]{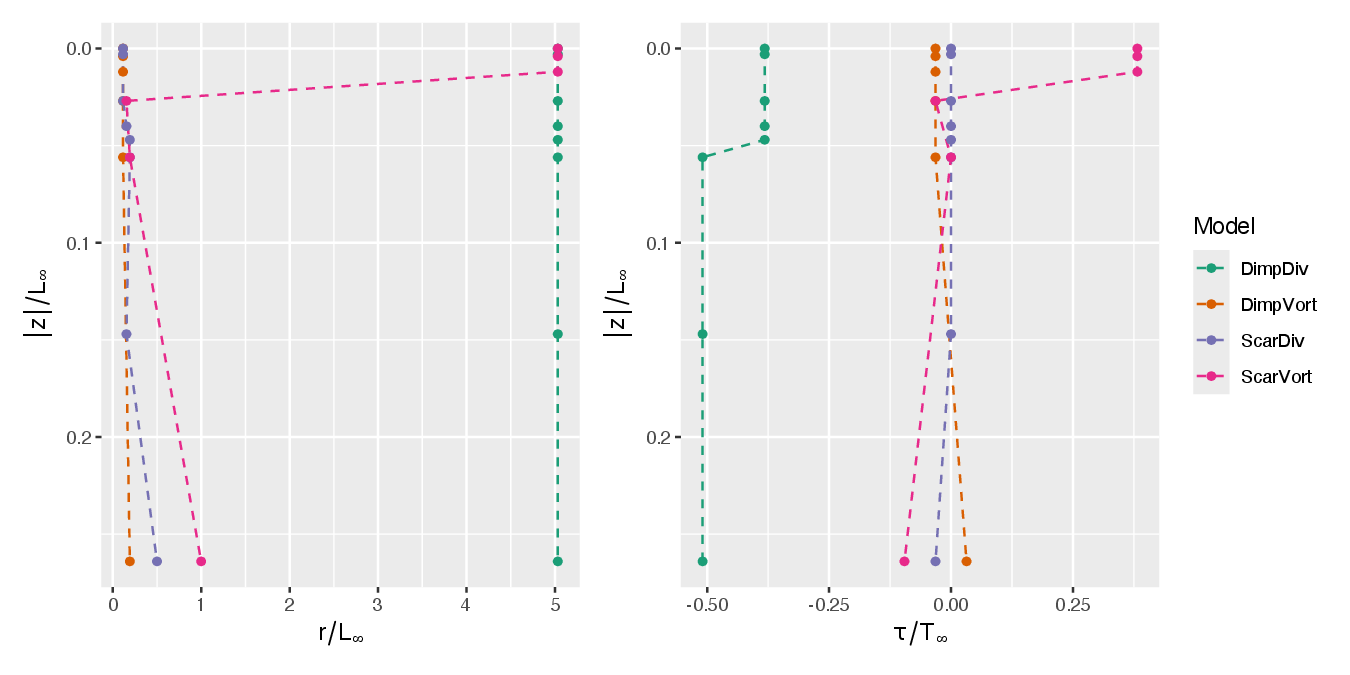}
            \put (33,12) {(a)}
            \put (80,12) {(b)}
    \end{overpic}
    \caption{
    Estimated paramater values $\hat{r}(z)$ (a) and $\hat{\tau}(z)$ (b) as functions of depth
    for Case 1 (similar plots for all cases may be found in supplementary materials); $r$ and $z$ are relative to integral scale $\Lint$. The points show $\hat{r}$, and $\hat{\tau}$ at the depths where we have done the analysis, while the dashed lines show clearer which points are connected. For the models and depths where the global model outperformed the local model we have used the parameter values for the global model (the optimal lag for the global model, and $\hat{r}(z)$ is set to the boundary value).
    }
    \label{fig:r_mle_varied_depth}
\end{figure}

The results presented so far showcase the performance of local models on the free surface, quantifying the qualitative narrative of the proximity of dimples and scars to regions of high variability of vertical vorticity and surface divergence, respectively. We next consider how the models perform \emph{below} the surface, by estimating model parameters plane by plane at increasing depth. The models estimate the connection between dimples and scars at the surface, with the vertical vorticity $\omega=\partial_x v-\partial_y u$ and horizontal divergence $\beta=\partial_x u+\partial_yv$ fields.

Figure \ref{fig:r_mle_varied_depth} depicts the optimal local support $r$ (panel a) and time shift $\tau$ (panel b) for a representative selection of depths, $z$, for all four local variance--models for Case 1. Consider first the orange and blue curves, representing the DimpVort and ScarDiv models for which a local model with small radius of support was optimal at the surface. 
The behaviour is consistent as the depth increases, with $r$ remaining small beneath the surface throughout the range we investigated (about a quarter of an integral length scale from the surface). 
The local model thus outperforms the global model also well beneath the surface. 
Similarly, the time shift is small, also below the surface, with a small increase at the greatest depth for ScarDiv. 
The DimpDiv model, on the other hand, is estimated to perform poorer than a global model of dimples and horizontal divergence, at the surface and at all depths tested below the surface. At all depths the likelihood is maximal at the largest admissible $r$, i.e. no local model outperforms the global one. The time shift remain negative below the surface, with a slight increase in absolute value at the largest depth.

Only the results for the ScarVort model change substantially with depth, from favouring a global model at the surface to preferring a local model as the depth increases. At a depth $|z|\approx 0.05 \Lint$, we see that the model has the same optimal local support $r$ and time shift $\tau$ as the ScarDiv and DimpVort models. The global preference of the model remains within the uppermost layer of the water column where viscous effects dominate the flow (the viscous layer; see \cite{shenSurfaceLayerFreesurface1999}). Below this, in the part of the surface-influenced region where surface blockage effects dominate (the blockage layer), the local model outperforms the global model. Note that even though the deepest point for which we report the correlation is much deeper than the viscous layer, it is still within the estimate of the extent of the large upwelling and downwelling flow events given by e.g., \cite{ruth2024}.

A main takeaway is that the two models that perform best locally at the surface---DimpVort and ScarDiv--continue to perform better with a local rather than a global model also at depths well into the blockage layer. For the DimpDiv and ScarVort models the global model is preferable at surface level, and while the latter changes to a local preference at depth, we hesitate to draw strong conclusions from this; scars generally lie in areas of strong intermittent flow and the preference is comparatively weak (not shown).


\section{Conclusions}
\label{sec:conclusions}

We have applied point process statistics, in order to study the degree of spatial locality when linking the turbulence of a free-surface turbulent flow to observable deformations at the surface, demonstrating the utility of this statistical method to turbulence research. 

Previous studies have shown strong linkage between surface features---to wit, dimples and scars---and sub-surface properties of the turbulent velocities, demonstrating strong correlation in \emph{time}, but not considering the spatial dimension; observed surface features were found to be good indicators that large, coherent sub-surface events were taking place, or recently did, somewhere in the vicinity. The flow features of special interest are long-lived surface-attached vortices and upwelling events where bulk water is brought to the surface. Using spatial statistics we can shed new light on what each surface feature type indicates not only concerning \emph{when} a sub-surface event happens, but also \emph{where}.

We have modelled dimples and scars as inhomogeneous Poisson point-processes, with intensity fields driven by the local variance of surface divergence, the vertical vorticity component, and the local mean of the signed surface divergence field. This enabled us to overcome the limitation imposed by spatially averaged statistics (which limit the analysis to temporal correlations), resulting in a quantification of the spatio-temporal relation between surface feature types---dimples and scars---and the relevant velocity derived flow fields---surface divergence and vertical vorticity.

The parameter fitting to our six models---four local-variance models and two local mean models with isolated signs--- demonstrate that dimples and scars are good proxies of spatial averages of surface divergence and vertical vorticity as they vary in time. 
However, the spatial dependencies between imprint type and field type are very different. 

Our results from the four local-variance models support spatially non-local---or `global'---relationships between scars and vertical vorticity at the surface whereas the strong correlation between the prevalence of dimples and vertical vorticity are highly local, predicting extrema in vertical vorticity to have dimples in their immediate vicinity---no surprise since dimples are well known to be imprints of surface-attached `bathtub' vortices, but a demonstration of the method. In contrast, scars appear a little before their concomitant surge in vertical vorticity, and there is next to no vertical vorticity exactly at the scar location (see \cite{aarnes2025}).
Less obvious is the highly local correlation between scars and surges in surface divergence, also essentially instantaneous. Centres of scars are found to lie along the perimeter of upwelling boils, with downwelling regions outside of them. Likewise the models suggest a global relationship between dimples and surface divergence, the correlation reported by \citet{babiker2023}.
There is thus a complementarity between dimples and scars, with respect to their local and global connections to surface divergence and vorticity. 

The correlation between dimples and extrema of surface divergence $\beta$ was further elucidated by considering areas of positive and negative $\beta$, separately. We devise two new local-mean models, one for each sign, and unlike the model for the unsigned $\beta^2$, both models have strong local behaviour, but in a sense opposite. For positive divergence---corresponding to upwelling---a local model effectively predicts where and when dimples are \emph{not} present, while still an indicator of dimple prevalence globally. Oppositely, negative-divergence indicates locally increased expected density of dimples. The result is intuitive when bearing in mind that dimples are imprints of vortices which move with the flow, which always points away from upwellings towards downwellings (convergence zones). The repulsion of vortices by upwellings is stronger and more immediate than the attraction to downwelling areas, resulting in a global model is best when only the squared value of $\beta$ is used, i.e., the DimpDiv model.

We further demonstrated how these local relationships between the surface deformations and turbulent fields change with depth, with the local relationship between scars and horizontal divergence gradually becoming less localised whereas the local relationship between dimples and vertical vorticity stays strongly local at all depths considered, all the way to the bottom of the blockage layer.

Finally, we wish to emphasize that the flows we considered are all in the region characterized by Brocchini and Peregrine \cite{brocchini2001a} as weak turbulence with only small surface deformations, and applications to other flow regimes should be done with care.

\backmatter

\section{Author contributions}
D.R.K.\ implemented and performed the statistical analysis. J.R.A.\ and O.M.B.\ prepared the data used in the analyses. I.S.\ and S.Å.E.\ supervised the project and contributed to discussions. D.R.K., J.R.A., and O.M.B.\ wrote the manuscript with edits and contributions from all authors.

\section{Data and Code Availability}
An implementation of the methods described here has been made available on GitHub: \hyperlink{https://github.com/danielKSt/codeSimData}{https://github.com/danielKSt/codeSimData}. The data we use is hosted on DataverseNO, see \cite{babiker23data,aarnes25data}.

\bmhead{Acknowledgements}
DNS data was generously shared with us by Prof.\ Lian Shen and Dr.\ Anqing Xuan at the University of Minnesota. We benefitted from discussion and input from Prof.\ Geir-Arne Fuglstad. AI was used to aid implementation and coding for analysis.
The research was co-funded by the Research Council of Norway (\emph{iMOD}, grant 325114) and the European Union (ERC CoG, \emph{WaTurSheD}).
Views and opinions expressed are, however, those of the authors only and do not necessarily reflect those of the European Union or the European Research Council. Neither the European Union nor the granting authority can be held responsible for them.

\appendix

\section{Different covariates and model selection}
\label{app:model_selection}
The results of Section \ref{sec:results_surface_variance} (where the up- and downwellings have not been isolated) use the local variance as the selected local flow property. However, other reasonable options exist.
In this appendix, we investigate how different choices of covariates perform in comparison to the local variance. We also look at the likelihood surfaces to investigate if they exhibit different behaviour compared to the local variance model.

\begin{figure}
    \centering
    \begin{overpic}[width=\textwidth]{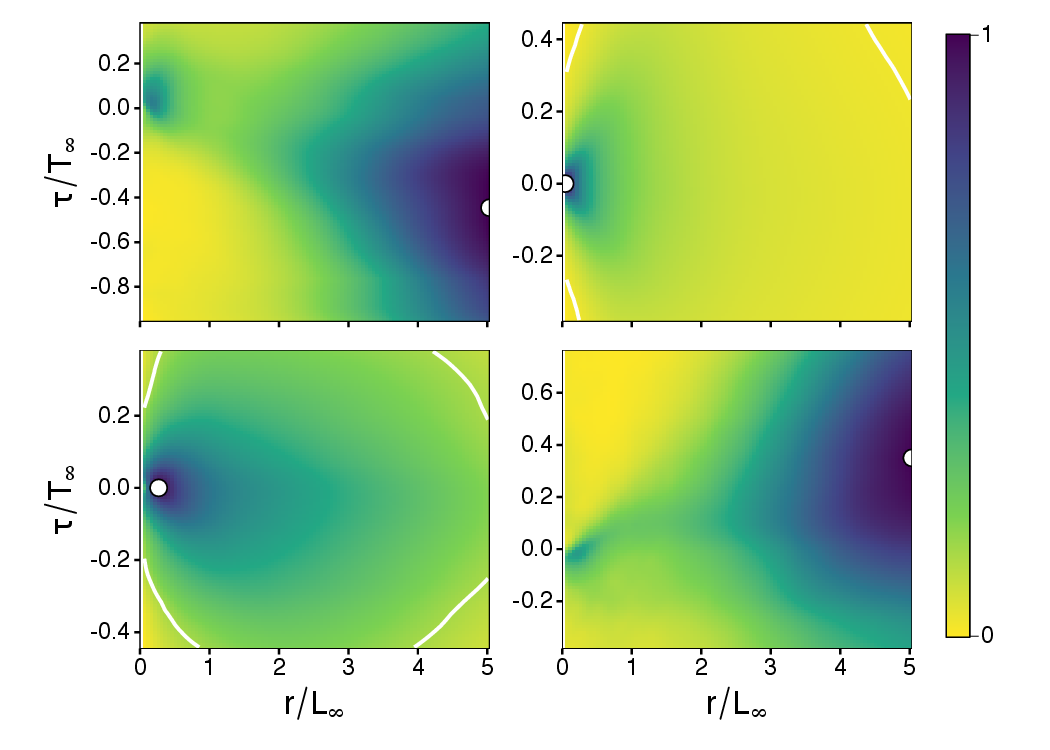}
            \put (15,65) {(a)}
            \put (57.5,65) {(b)}
            \put (15,32) {(c)}
            \put (57.5,32) {(d)}
    \end{overpic}
    \caption{Normalized profile likelihood for the models (a) DimpDiv, (b) DimpVort, (c) ScarDiv and (d) ScarVort, for Case 1 (see Table \ref{tab:flopProp}), using local mean square as explanatory variable. Larger values indicate better performance. Spatial support $r$ and temporal shift $\tau$ are normalized with the integral length and time scales of the turbulence, respectively.
    }
    \label{fig:likelihood_surface_meanSquare}
\end{figure}
A natural candidate alternative field is the local mean square,
\begin{equation}
    \label{eq:local_mean_square}
    \overline{\sfield^2_r}(x,y,t) = \frac{1}{\pi r^2}\int_{B_r(x,y)}\sfield^2(\xi,\eta,t+\tau)d\xi d\eta,
\end{equation} 
which is strictly positive and can be inserted into the power law model without any changes to the fitting approach. (Clearly, $\overline{\sfield^2_r}=\sigma_\phi^2+\overline{\phi_r}^2$; see Eq.~\ref{eq:sigphi})

Figure \ref{fig:likelihood_surface_meanSquare} depicts the profile likelihoods obtained for the local mean square.
We observe that these show a similar overall picture compared to the profile likelihoods for local variance in Figure \ref{fig:likelihood_surface}, with similar overall shapes and scales. 
In particular, the complementarity of dimples and scars that we found with the local variance has not changed. 
One observable difference is that the local maximum for DimpVort appears sharper and at a shorter radius than for the local variance model. The difference stems from the local variance at very short radii showing the edge of dimples, rather than the centre, while the local mean square with short radii shows directly where the centroid of a dimple is.

It is then also of interest to compare whether local variance or local mean square has greater explanatory power. To this end we provide a table of change in log-likelihood from the baseline model of a constant intensity, to the relevant model with the best choice of parameters. These values are presented in Table \ref{tab:local_flow_properties_comparisons}. We see then that for the DimpVort model, the local mean square outperforms the local variance, yet the opposite is true for the ScarDiv model.

\begin{table}
\centering
\begin{tabular}{|l|ccc|ccc|}
    \hline
     & \multicolumn{3}{c|}{Surface divergence} & \multicolumn{3}{c|}{Vertical vorticity} \\
    \hline
     & $\sigma_\phi^2$ & $\overline{\sfield^2_r}$ & Global & $\sigma_\phi^2$ & $\overline{\sfield^2_r}$ & Global \\
    \hline
    Dimples & 292.1 & 292.4 & \textbf{312.2} & 4719.4 & \textbf{7879.6} & 334.3 \\
    Scars & \textbf{3623.6} & 2286.0 & 447.8 & 348.3 & 348.4 & \textbf{354.2} \\
    \hline
\end{tabular} 
\caption{Comparison of performance for various local flow properties. The performance is measured in $\Delta \log L$, the change in log-likelihood from the null model where the intensity $\nu$ is constant in space and time.}
\label{tab:local_flow_properties_comparisons}
\end{table}

\section{Model diagnostics}
\label{app:diagnostics}
Throughout this paper, we have utilized the Poisson point process model. In the present appendix, we investigate how well the model assumptions hold for the point patterns we study, and present uncertainty estimates for the model.

\subsection{Clustering diagnostics}
\label{app:clustering}
The Poisson point process model requires independent point counts between disjoint areas, given the intensity. In point process literature, a common way of measuring interaction in point patterns is the pair correlation function (pcf); see chapter 4.3 in \cite{Illian2008StatisticalAA} for a detailed introduction to this and other measures of clustering. 

The pcf, denoted $g(h)$, measures if there is a change in the probability of more points at a distance $h$ from a point in the point pattern. $g(h)>1$ indicates that the probability increases (clustering at distance $h$), and $g(h)<1$ indicates that the probability decreases (repulsion at distance $h$). In point patterns where there is no interaction between points the pcf is identically 1. 

The pcf of the fitted model, that is, the pcf of 
a Poisson point pattern with an intensity following one of the models in Table \ref{tab:Models}, can then be compared with the estimated pcf of the observed surface features. If the clustering/repulsion in the point pattern is explained well by the intensity and the Poisson point pattern model, then the pcf of the fitted model and the estimated pcf should be close. We denote the pcf estimated from the observed data as $\hat{g}_{(s/d)}(h)$ and the pcf of the fitted model as $g_{(s/d)}(h)$, with the subscript $s$ and $d$ indicating if this is the pcf for scars or dimples.

\begin{figure}
    \centering
    \includegraphics[width=0.9\linewidth]{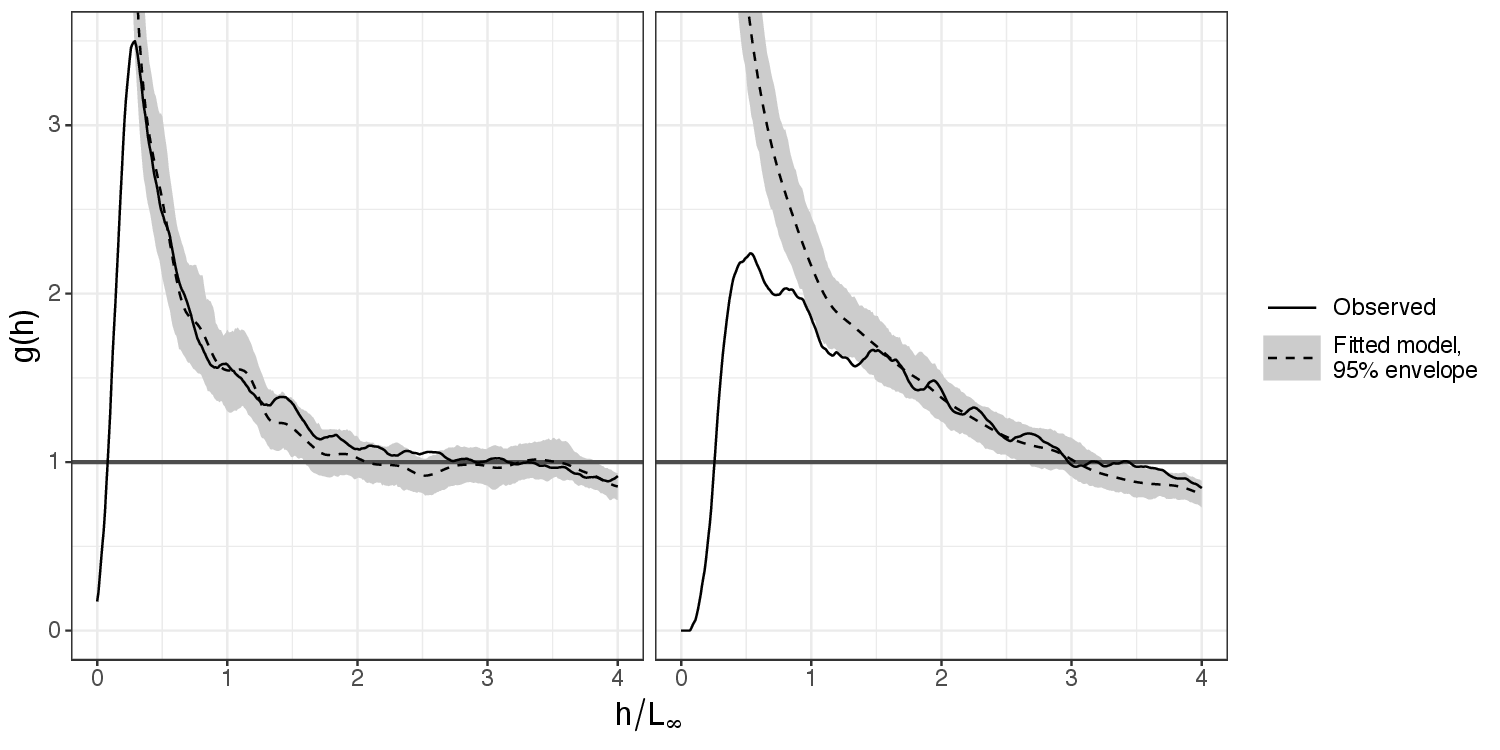}
    \caption{The estimated pcf of the observed data in Case 1. The left panel shows results for the DimpVort model(using local mean square as local flow property), and the right panel shows results for the ScarDiv (using local variance as local flow property). The solid line is the pcf estimated from the observed surface features. The dashed line gives an estimate of the model pcf for the fitted model, while the shaded area displays an estimated 95\% confidence interval for the pcf of the fitted model.}
    \label{fig:pcf}
\end{figure}

Figure \ref{fig:pcf} displays $\hat{g}_{(s/d)}(h)$, with dimples on the left and scars on the right, along with $g_{(s/d)}(h)$. For both dimples and scars, $\hat{g}_{(s/d)}(h)$ is near zero at $h=0$, indicating repulsion at short distances in the observed dimples and scars. As $h$ increases, $\hat{g}_{(s/d)}(h)$ grows larger than $1$, indicating clustering at what we may call a medium range. At large range we see $\hat{g}_{(s/d)}(h)$ is near one, i.e., no interaction between dimples or between scars. For medium and large ranges $\hat{g}_{(s/d)}(h)$ matches $g_{(s/d)}(h)$ very well, but $g_{(s/d)}(h)$ does not exhibit the short range repulsion found in the observed data. 

The mismatch between $g_{(s/d)}(h)$ and $\hat{g}_{(s/d)}(h)$ for small $h$ can be explained by noting that both dimples and scars occupy physical space (or simulated surface area), limiting the closest possible distance between centroids. In addition, the spatial simulation resolution imposes a hard limit on the closeness of two identified objects. The effect of these physical and computational constraints come out as a repulsion effect in the pcf: Since two dimples or scars cannot occupy the same space or lie arbitrary close to each other, they ``repel" one another. Scars are usually longer than dimples, which causes the apparent repulsion to reach larger radii in $\hat{g}_{(s)}(h)$ than in $\hat{g}_{(d)}(h)$.
The Poisson pattern model does not allow for interaction between points given the intensity, thus this model is not able to capture the observed core repulsion in $\hat{g}_{(s/d)}(h)$. 
There are models in the point process literature that allow for short range repulsion and medium range clustering as seen in $\hat{g}_{(s/d)}(h)$ (e.g., Mat\'{e}rn thinned Cox processes \cite{Andersen2016MatrnTC}), outside the scope of this paper.

\subsection{Snapshot independence}
\label{app:snapshots_independence}
Our likelihood calculations presented in Sec.~\ref{sec:results} assumed independence between subsequent snapshots, an assumption we shall test here. In Table \ref{tab:snapshot_overview}, we present a list of snapshot counts along with the minimal and maximal lags used for the parameter search space, $\lagmin$ and $\lagmax$. The auto-correlation of the structure count using one integral time scale to separate the snapshots is $0.52$ for dimples, and $0.29$ for scars. The higher value for dimples reflects the fact that surface-attached vortices decay slowly and can persist considerably longer than any bulk turbulence timescale.

While a lower value would be ideal, we need to compromise slightly due to intermittency and the limited amount of data; if strong events fall between timesteps and are not captured, inferred statistics may not be fully representative.

To determine the effect of snapshot independence or lack thereof, we performed the analysis using a time step of two integral time scales rather than one between snapshots. The results (not shown) yield nearly identical values for the radii and lag of the local models. Thus, we conclude that the analysis is not sensitive to the choice of using a time separation of one integral time scale.

\begin{table}[]
\centering
\begin{tabular}{|c|c|cc|}
    \hline
    Case & $n$ & $\lagmin/\Tint$ & $\lagmax/\Tint$ \\
    \hline
    1 & 78 & $-1.15$ & $0.76$ \\
    2 & 100 & $-1.17$ & $0.75$ \\
    3 & 101 & $-1.17$ & $0.76$ \\
    4 & 97 & $-1.18$ & $0.73$ \\
    5 & 93 & $-1.14$ & $0.79$ \\
    6 & 95 & $-1.13$ & $0.78$ \\
    \hline
\end{tabular}
\caption{Snapshot count $n$, along with the timeshift limits $\lagmin$, and $\lagmax$, for the six turbulence simulation cases.}
\label{tab:snapshot_overview}
\end{table}

\subsection{Quantification of uncertainty}
\label{app:uncertainty}

\begin{figure}
    \centering
    \includegraphics[width=0.95\linewidth]{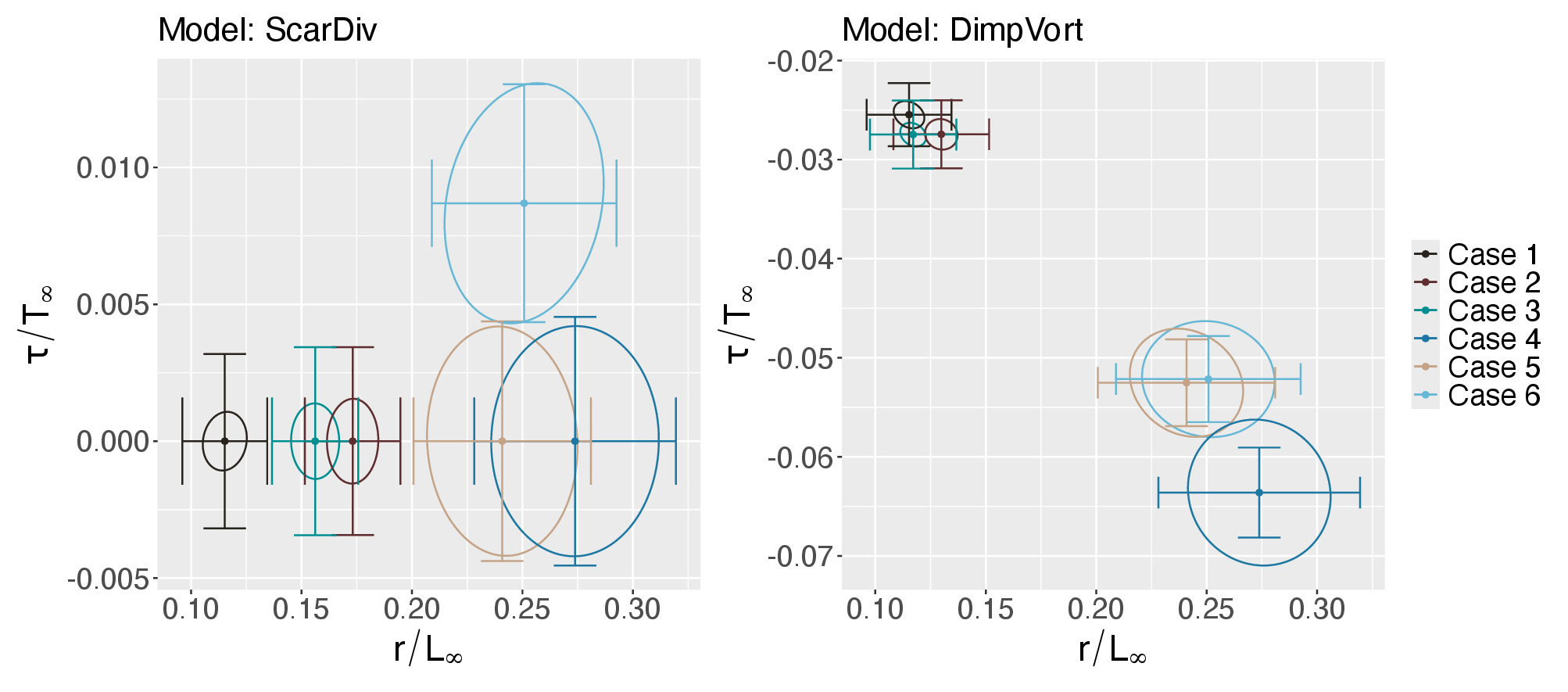}
    \caption{Estimated parameters (dots), 95\% confidence intervals (ellipses), and simulation discretization scales (errorbars) for all six cases for models ScarDiv and DimpVort.
    }
    \label{fig:confidence}
\end{figure}

The results in Fig.~\ref{fig:likelihood_surface} are based on parameter estimates for $r$ and $\tau$. When using maximum likelihood estimation to estimate parameters, the standard approach to uncertainty quantification is to use the inverse of the Hessian (estimated by finite differences) of the log-likelihood at the ML estimate, which under certain asymptotic assumptions can be used to construct an approximation of a $95\%$ confidence region, see Chapter 10.1 of \cite{casellaBergerStatInf}.
We use this approach for the two models where the local model outperforms the global model, that is, for ScarDiv and DimpVort. The results are plotted in Fig.~\ref{fig:confidence}. The figure depicts the optimised parameters (dots) with $95\%$ confidence regions (ellipses) for all six flow cases for each of the two models, and compares these to the spatial and temporal resolution (horizontal and vertical error bars, respectively) for each case. Note that the resolution of the simulations is not incorporated into the likelihood. This means that the uncertainty can possibly be smaller than the resolution.

In all cases, the confidence regions of $r$ and $\tau$ are of the same order as the spatial and temporal resolutions of the DNS simulations, respectively. Indeed, the confidence regions are smaller than the resolution for the cases with high turbulence levels (case 1--3) and about the same as the resolution for the cases with low turbulence level (case 4--6).
The main take-away is that our models give high confidence that a local and near-instantaneous model with local support of less than half an integral scale's radius far outperforms a global model when using scars to identify high divergence or dimples to identify vorticity peaks.

We note that not all of the assumptions of the likelihood model are fully satisfied. The snapshots are not completely independent (see Appendix \ref{app:snapshots_independence}), and there is some repulsion at shorter inter-point distances (see Appendix \ref{app:clustering}). Therefore, the confidence regions shown in the figure are likely to be somewhat too optimistic.

\section{Detected dimples}
\label{app:detected_dimps}

In Section \ref{sec:datasets}, we identify dimples via their attached vortices, found from the velocity field using the $\lambda_2$ criterion by \citet{jeongIdentificationVortex1995}. In potential applications (e.g., remote sensing applications), the velocity field would not be available and dimples must be identified from surface information only. In the following section, we present the results we obtain by repeating the analysis from the main text on dimples detected by the wavelet-based computer vision method described in \cite{babiker2023}.

\begin{figure}
    \centering
    \begin{overpic}[width=\textwidth]{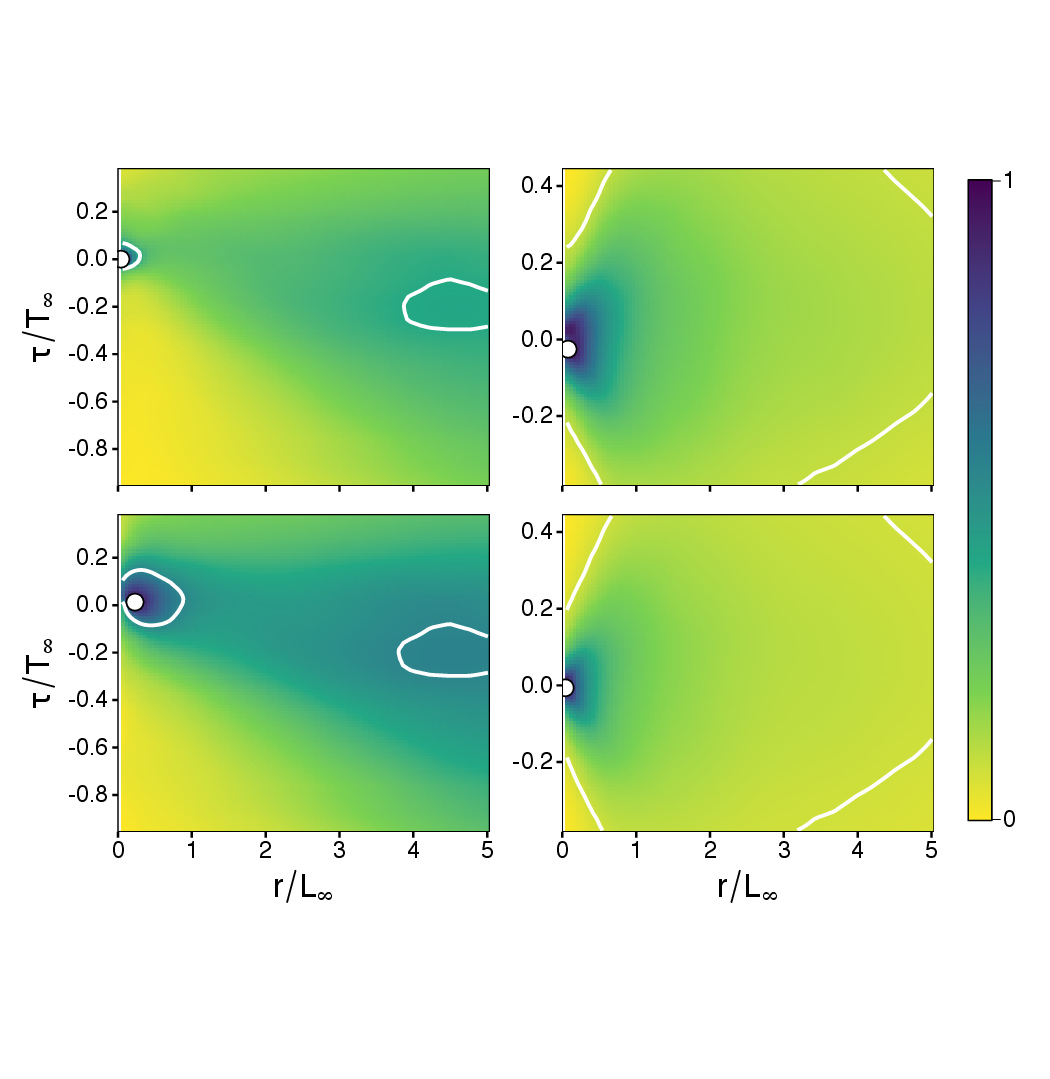}
    \put (15,15) {(a)}
    \put (80,15) {(b)}
    \end{overpic}
    \caption{Normalized profile likelihood using dimples detected with the computer vision method of \cite{babiker2023} rather than vortices found from the velocity field, for Case 1 (see Table \ref{tab:flopProp}), with models DimpDiv (a) and DimpVort (b) using local variance.
    }
    \label{fig:detected_dimples_likSurf}
\end{figure}

Figure \ref{fig:detected_dimples_likSurf} shows the resulting profile likelihoods for the DimpDiv and DimpVort models, corresponding to Fig.~\ref{fig:likelihood_surface}(a) and (b).
The shape of the profile likelihood for DimpVort is virtually identical to that of Figure \ref{fig:likelihood_surface}(b).
There is a greater difference when comparing the profile likelihood for DimpDiv: When using detected  dimples, the results indicate a local model preference with extremely small radius and no lag, contrary to what was observed when using vortices computed from the velocity field. The local maximum in the likelihood is so strong that the increasing values for higher $r$ and negative lags---very clear in Fig.~\ref{fig:likelihood_surface}(a)---is somewhat obscured after rescaling values to the unit interval.

A likely reason for the difference is false positives in the dimple detection algorithm.
As stated in \citet{babiker2023}, 5--10\% of the detected dimples are false positives, further investigated in \citet[][ch. 3]{babikerPhD} as mainly arising when scars are mislabelled as dimples. 
Referring to the maximum log-likelihood in Table \ref{tab:detected_dimples_log_lik_score_comariasons}, we observe that while the local variance of the surface divergence now outperforms the global model, the vertical vorticity still has a much stronger local coupling. Moreover, comparing these to the local and global values for ScarDiv in Table \ref{tab:local_flow_properties_comparisons} supports the claim that any scars misidentified as dimples and used in the DimpDiv model, will have a large effect towards skewing the log-likelihood towards a local model.

Indeed, our analysis in Sec.~\ref{sec:DivUpDivDown} showed that a highly local model for dimples and surface divergence would predict either where dimples most likely \emph{are} or where they most likely \emph{are not}, depending on whether a surge in $|\beta|$ is a downwelling or an upwelling, respectively, so a local model with very small compact support is a poor choice. 
A similar extremum was observed for very small radii also for the `regular' DimpDiv in Fig.~\ref{fig:likelihood_surface}(a), and is much strengthened here.

While it is true that the detection method of Ref.~\cite{babiker2023} has room for improvement and reducing the percentage of false positives further would help isolate a ``true'' optimal $r$, a prudent interpretation of Fig.~\ref{fig:detected_dimples_likSurf} is already to recommend a global model for DimpDiv.

\begin{table}
\centering
\begin{tabular}{|l|ccc|ccc|}
    \hline
     & \multicolumn{3}{c|}{Surface divergence} & \multicolumn{3}{c|}{Vertical vorticity} \\
    \hline
     & $\sigma_\phi^2$ & $\overline{\sfield^2_r}$ & Global & $\sigma_\phi^2$ & $\overline{\sfield^2_r}$ & Global \\
    \hline
    Detected dimples & \textbf{999.7} & 711.4 & 382.0 & 2753.0 & \textbf{4116.4} & 397.2 \\
    \hline
\end{tabular}
\caption{Comparison of performance for various local flow properties against the detected dimples. The performance is measured in $\Delta \log L$, the change in log-likelihood from the null model where the intensity $\nu$ is constant in space and time.}
\label{tab:detected_dimples_log_lik_score_comariasons}
\end{table}

\bibliographystyle{bst/sn-mathphys-num}
\bibliography{spat}

@article{aarnes2025,
   author = {Aarnes, J{\o}rgen R. and Babiker, Omer M. and Xuan, Anqing and Shen, Lian and Ellingsen, Simen {\AA}.},
   title = {Vortex structures under dimples and scars in turbulent free-surface flows},
   journal = {J. Fluid Mech.},
   volume = {1007},
    pages = {A38},
   ISSN = {0022-1120},
   DOI = {10.1017/jfm.2025.72},
   year = {2025},
   type = {Journal Article}
}

@article{moen2026,
  title = {Mapping surface height dynamics to subsurface flow physics in free-surface turbulent flow using a shallow recurrent decoder},
  author = {Moen, Kristoffer S. and Aarnes, J\o{}rgen R. and Ellingsen, Simen {\AA}. and Kutz, J. Nathan},
  journal = {Phys. Rev. Fluids},
  volume = {11},
  issue = {8},
  pages = {084601},
  numpages = {33},
  year = {2026},
  month = {Aug},
  publisher = {American Physical Society},
  doi = {10.1103/3q12-ylb6},
  url_ = {https://link.aps.org/doi/10.1103/3q12-ylb6}
}

@article{janzen2010,
   author = {Janzen, Johannes G and Herlina, H and Jirka, Gerhard H and Schulz, Harry E and Gulliver, John S},
   title = {Estimation of mass transfer velocity based on measured turbulence parameters},
   journal = {AIChE journal},
   volume = {56},
   number = {8},
   pages = {2005-2017},
   ISSN = {0001-1541},
   year = {2010},
doi = {10.1002/aic.12123},
   type = {Journal Article}
}

@misc{Ruth2026,
   author = {Ruth, Daniel J and Coletti, Filippo},
   title = {Free-surface curvature and its relation to subsurface turbulence},
   note = {arXiv preprint},
   url = {https://arxiv.org/abs/2608.04687},
   year = {2026}
}

@misc{babiker23data,
author = {Babiker, O. M. and Bjerkeb{\ae}k, I. and Xuan, A. and Shen, L. and Ellingsen, S. {\AA}.},
publisher = {DataverseNO},
title = {Supporting data for: {V}ortex imprints on a free surface as proxy for surface divergence.},
year = {2023},
version = {DRAFT VERSION},
_note = {{URL}: DOI to be added in proofs},
doi = {10.18710/ZOJHGJ},
url = {10.18710/ZOJHGJ}
}

@misc{aarnes25data,
author = {Aarnes, J{\o}rgen R. and Babiker, Omer M. and Xuan, Anqing and Shen, Lian and Ellingsen, Simen {\AA}.},
publisher = {DataverseNO},
title = {Replication data for: “Vortex structures under dimples and scars in turbulent free-surface flows” Dataverse, Part 1},
year = {2025},
version = {1.1},
doi = {10.18710/XQ81WH},
url = {10.18710/XQ81WH}
}

@article{babiker2023,
  title = {Vortex Imprints on a Free Surface as Proxy for Surface Divergence},
  author = {Babiker, O. M. and Bjerkeb{\ae}k, I. and Xuan, A. and Shen, L. and Ellingsen, S. {\AA}.},
  year = {2023},
  journal = {J. Fluid Mech.},
   DOI = {10.1017/jfm.2023.370},
  volume = {964},
  pages = {R2}
}

@article{babiker2026,
  title = {Experimental investigation relating free-surface features to subsurface turbulence},
  author = {Babiker, Omer M. and Aarnes, J{\o}rgen R. and Semati, Ali and Ferran, Am\'elie and Tee, Yi Hui and Hearst, R. Jason and Ellingsen, Simen {\AA}.},
  journal = {Physical Review Fluids},
  volume = {11},
  issue = {5},
  pages = {054802},
  numpages = {32},
  year = {2026},
  month = {May},
  publisher = {American Physical Society},
  doi = {10.1103/bmx7-2z3h},
  url = {https://link.aps.org/doi/10.1103/bmx7-2z3h}
}

@phdthesis{babikerPhD,
  title={Turbulent imprints on the free surface},
  author={Babiker, Omer M.},
  year={2026},
  school={Norwegian University of Science and Technology}
}

@article{banerjee1994,
   author = {Banerjee, S.},
   title = {Upwellings, Downdrafts, and Whirlpools: Dominant Structures in Free Surface Turbulence},
    journal = {Appl. Mech. Rev.},   
   volume = {47},
   number = {6S},
   pages = {S166-S172},
   ISSN = {0003-6900},
   DOI = {10.1115/1.3124398},
   year = {1994},
   type = {Journal Article},
}

@article{banerjee2004,
  title = {Surface Divergence Models for Scalar Exchange between Turbulent Streams},
  year = {2004},
  journal = {Int. J. Multiph. Flow},
  volume = {30},
  number = {7},
  pages = {963--977},
  issn = {0301-9322},
  DOI = {10.1016/j.ijmultiphaseflow.2004.05.004},
  author = {Banerjee, S. and Lakehal, D. and Fulgosi, M.}
}

@article{brocchini2001a,
  title = {The Dynamics of Strong Turbulence at Free Surfaces. {Part} 1. {Description}},
  author = {Brocchini, M. and Peregrine, D. H.},
  year = {2001},
  month = dec,
  journal = {J. Fluid Mech.},
  doi={10.1017/S0022112001006012},
  volume = {449},
  pages = {225--254},
}

@article{guoInteractionDeformableFree2010,
  title = {Interaction of a Deformable Free Surface with Statistically Steady Homogeneous Turbulence},
  author = {Guo, Xin and Shen, Lian},
  year = {2010},
  DOI = {10.1017/S0022112010001539},
  ISSN = {0022-1120},
  month = sep,
  journal = {J. Fluid Mech.},
  volume = {658},
  pages = {33--62}
}

@article{jeongIdentificationVortex1995,
  title = {On the Identification of a Vortex},
  author = {Jeong, J. and Hussain, F.},
  year = {1995},
  month = {feb},
  journal = {J. Fluid Mech.},
  doi = {10.1017/S0022112095000462},
  volume = {285},
  pages = {69--94}
}

@article{kermani2009,
   author = {Kermani, A. and Shen, L.},
   title = {Surface age of surface renewal in turbulent interfacial transport},
   journal = {Geophys, Res. Letters}, _journal = {geophysical Research Letters},
   volume = {36},
   number = {10},
   pages = {L10605–n/a},
   ISSN = {0094-8276},
   DOI = {10.1029/2008GL037050},
   year = {2009},
   type = {Journal Article}
}

@article{khakpour2012,
  title = {Coherent Vortical Structures Responsible for Strong Flux of Scalar at Free Surface},
  year = {2012},
  journal = {Int. J. Heat Mass Tran.},
  volume = {55},
  number = {19},
  pages = {5157--5170},
  issn = {0017-9310},
  DOI = {10.1016/j.ijheatmasstransfer.2012.05.017},
  author = {Khakpour, H. R. and Igusa, T. and Shen, L.}
}

@article{longuet-higgins96,
   author = {Longuet-Higgins, M. S.},
   title = {Surface manifestations of turbulent flow},
   journal = {J. Fluid Mech.},
   doi={10.1017/S0022112096001371},
   year = {1996},
   volume = {308},
   pages = {15-29}
}

@article{McKenna2004,
   author = {McKenna, Sean P and McGillis, Wade R},
   title = {The role of free-surface turbulence and surfactants in air–water gas transfer},
   journal = {Int. J. Heat Mass Tran.},
   volume = {47},
   number = {3},
   pages = {539-553},
   ISSN = {0017-9310},
   DOI = {10.1016/j.ijheatmasstransfer.2003.06.001},
   year = {2004},
   type = {Journal Article}
}

@article{ruth2024,
   author = {Ruth, Daniel J. and Coletti, Filippo},
   title = {Structure and energy transfer in homogeneous turbulence below a free surface},
   journal = {J. Fluid Mech},
   volume = {1001},
   ISSN = {0022-1120},
   DOI = {10.1017/jfm.2024.1017},
   year = {2024},
   type = {Journal Article}
}

@article{shenSurfaceLayerFreesurface1999,
  title = {The Surface Layer for Free-Surface Turbulent Flows},
  author = {Shen, Lian and Zhang, Xiang and Yue, Dick K. P. and Triantafyllou, George S.},
  year = {1999},
  month = may,
  journal = {J. Fluid Mech.},
  volume = {386},
  pages = {167--212},
  issn = {0022-1120, 1469-7645},
  doi = {10.1017/S0022112099004590},
  urldate = {2023-03-21},
  langid = {english}
}

@article{pinelli2022,
   author = {Pinelli, Michele and Herlina, Herlina and Wissink, Jan G and Uhlmann, Markus},
   title = {Direct numerical simulation of turbulent mass transfer at the surface of an open channel flow},
   journal = {J. Fluid Mech.}, _journal = {Journal of Fluid Mechanics},
   volume = {933},
   pages = {A49},
   ISSN = {0022-1120},
   year = {2022},
   doi={10.1017/jfm.2021.1080},
   type = {Journal Article}
}

@article{turney2013,
  title = {Air\textendash Water Gas Transfer and near-Surface Motions},
  author = {Turney, D. E. and Banerjee, S.},
  year = {2013},
  month = oct,
  journal = {J. Fluid Mech.},
  volume = {733},
  pages = {588--624},
  publisher = {Cambridge University Press},
  issn = {0022-1120, 1469-7645},
  doi = {10.1017/jfm.2013.435}
}

@article{turney2005,
   author = {Turney, Damon E and Smith, Walter C and Banerjee, Sanjoy},
   title = {A measure of near‐surface fluid motions that predicts air‐water gas transfer in a wide range of conditions},
   journal = {Geophys. Res. Lett.},
   volume = {32},
   pages={L04607},
   DOI = {10.1029/2004GL021671},
   number = {4},
   ISSN = {0094-8276},
   year = {2005},
   type = {Journal Article}
}

@article{xuanConservativeSchemeSimulation2019,
  title = {A Conservative Scheme for Simulation of Free-Surface Turbulent and Wave Flows},
  author = {Xuan, Anqing and Shen, Lian},
  year = {2019},
  month = feb,
  journal = {J. Comput. Phys.},
  volume = {378},
  pages = {18--43},
  issn = {0021-9991},
  doi = {10.1016/j.jcp.2018.10.046},
  urldate = {2023-03-28}
}

@article{rosalesLinearForcingNumerical2005,
  title = {Linear Forcing in Numerical Simulations of Isotropic Turbulence: {{Physical}} Space Implementations and Convergence Properties},
  shorttitle = {Linear Forcing in Numerical Simulations of Isotropic Turbulence},
  author = {Rosales, Carlos and Meneveau, Charles},
  year = {2005},
  month = sep,
  journal = {Phys. Fluids},
  volume = {17},
  number = {9},
  pages = {095106},
  publisher = {{American Institute of Physics}},
  issn = {1070-6631},
  doi = {10.1063/1.2047568},
  urldate = {2023-03-31},
}

@book{Illian2008StatisticalAA,
    title = {Statistical Analysis and Modelling of Spatial Point Patterns},
    author = {Illian, J. B. and Penttinen, A. and Stoyan, H. and Stoyan, D.},
    year = {2007},
    publisher = {John Wiley \& Sons, Ltd},
    isbn = {9780470725160},
    doi = {10.1002/9780470725160},
    url = {https://onlinelibrary.wiley.com/doi/abs/10.1002/9780470725160}
}

@book{spatstat,
    author = {Baddeley, A. and Rubak, E. and Turner, R.},
    title = {Spatial Point Patterns: Methodology and Applications with R (1st ed.)},
    publisher = {Chapman and Hall/CRC},
    year = {2015},
    doi = {10.1201/b19708}
}

@book{GLM_fahrmeir,
  title={Regression: Models, Methods and Applications},
  author={Fahrmeir,L and Kneib, T and Lang, S and Marx, B. D.},
  publisher = {Springer New York, NY},
  year={2013},
  doi = {10.1007/978-3-662-63882-8},
  url={https://link.springer.com/book/10.1007/978-3-662-63882-8}
}

@book{GLM_mcCullagh,
  title={Generalized Linear Models},
  author={McCullagh, P. and Nelder, J.A.},
  publisher = {Chapmann \& Hall / CRC},
  year={1989},
  doi = {10.1201/9780203753736},
  url={https://www.taylorfrancis.com/books/mono/10.1201/9780203753736/generalized-linear-models-mccullagh}
}

@book{casellaBergerStatInf,
  title={Statistical Inference},
  author={George Casella and Roger L. Berger},
  year={1990},
  publisher = {Chapmann \& Hall / CRC},
  url={https://api.semanticscholar.org/CorpusID:265812986}
}

@article{Andersen2016MatrnTC,
	title={Mat{\'e}rn thinned Cox processes},
	author={Ina Trolle Andersen and Ute Hahn},
	journal={spatial statistics},
	year={2016},
	volume={15},
	pages={1-21},
	url={https://api.semanticscholar.org/CorpusID:123188733}
}

@article{brocchini2001dynamics,
  title={The dynamics of strong turbulence at free surfaces. Part 2. Free-surface boundary conditions},
  author={Brocchini, Maurizio and Peregrine, DH},
  journal={J. Fluid Mech.}, _journal = {Journal of Fluid Mechanics},
  volume={449},
  pages={255--290},
  doi={10.1017/S0022112001006012},
  year={2001},
  publisher={Cambridge University Press}
}

@article{qi2025small,
  title={Small-scale dynamics and structure of free-surface turbulence},
  author={Qi, Yinghe and Li, Yaxing and Coletti, Filippo},
  journal={J. Fluid Mech.}, _journal = {Journal of Fluid Mechanics},
  volume={1007},
  doi =  {10.1017/jfm.2025.139},
  pages={A3},
  year={2025},
  publisher={Cambridge University Press}
}

@article{wu2026localised,
  title={Localised inter-scale energy cascades in free-surface turbulence},
  author={Wu, Guotao and Li, Yaxing and Coletti, Filippo},
  journal={J. Fluid Mech.}, _journal = {Journal of Fluid Mechanics},
  volume={1026},
  doi =  {10.1017/jfm.2025.11071},
  pages={A51},
  year={2026},
  publisher={Cambridge University Press}
}

\end{document}